\documentclass[%
superscriptaddress,
preprint,
 amsmath,amssymb,
 aps,
]{revtex4-1}
\usepackage{setspace}
\usepackage{tabto}
\usepackage{xcolor}
\usepackage{algorithm}
\usepackage{algorithmicx}
\usepackage{algpseudocode}
\usepackage{qcircuit}
\usepackage{braket}
\usepackage[labelfont=bf,justification=justified,format=plain]{caption}
\usepackage{subcaption}
\usepackage[normalem]{ulem}

\usepackage{graphicx}% Include figure files
\usepackage{dcolumn}% Align table columns on decimal point
\usepackage{bm}% bold math
\usepackage{xcolor}
\usepackage{qcircuit}
\usepackage{braket}
\usepackage{float}
\usepackage{algorithm}
\usepackage{algorithmicx}
\usepackage{algpseudocode}
\usepackage{mathtools}
\usepackage{multirow}
\usepackage{hyperref}% add hypertext capabilities
\renewcommand{\figureautorefname}{Figure~\negthinspace}
\renewcommand{\equationautorefname}{Equation~\negthinspace}
\renewcommand{\tableautorefname}{Table~\negthinspace}
\renewcommand{\sectionautorefname}{Section~\negthinspace}
\renewcommand{\appendixautorefname}{Appendix~\negthinspace}
\newcommand{\algorithmautorefname}{Algorithm~\negthinspace}

\DeclareMathOperator*{\E}{\mathbb{E}}

\begin{document}

\preprint{BNL}

\title{Quantum Architecture Search via Continual Reinforcement Learning}% Force line breaks with \\

\author{Esther Ye}%
 \email{estherye@bu.edu}
\affiliation{%
 Department of Electrical and Computer Engineering, Boston University, Boston, MA 02215, USA
}%

\author{Samuel Yen-Chi Chen}
\email{ychen@bnl.gov}
\affiliation{%
 Computational Science Initiative, Brookhaven National Laboratory, Upton, NY 11973, USA
}%

% \collaboration{MUSO Collaboration}%\noaffiliation

%  \homepage{http://www.Second.institution.edu/~Charlie.Author}
% \affiliation{
%  Second institution and/or address\\
%  This line break forced% with \\
% }%
% \affiliation{
%  Third institution, the second for Charlie Author
% \affiliation{%
%  Authors' institution and/or address\\
%  This line break forced with \textbackslash\textbackslash
% }%

% \collaboration{CLEO Collaboration}%\noaffiliation

\date{\today}% It is always \today, today,

\begin{abstract}
% Quantum computing has been developing over the years and it proves to be an enthralling topic because it is not a trivial task to design quantum circuitry without expert knowledge. 
Quantum computing has promised significant improvement in solving difficult computational tasks over classical computers. 
%
%As such, recent advances in quantum computing hardware have drawn 
%significant
%attention to quantum algorithms and applications. 
%
Designing quantum circuits 
%architectures 
for practical use, however, is not a trivial objective and requires expert-level knowledge.
%
% However, designing quantum circuit architecture for practical use is not a trivial task and requires expert level knowledge. To make situation more complicated, available quantum computers are noisy, therefore the effect of noises need to be taken into account.
%
To aid this endeavor, this paper proposes a machine learning-based method to construct quantum circuit architectures. 
Previous works have demonstrated that classical deep reinforcement learning (DRL) algorithms can successfully construct quantum circuit architectures without encoded physics knowledge. 
However, these DRL-based works are not generalizable to settings with changing device noises, thus requiring  
%significant
considerable amounts of training resources to keep the RL models up-to-date.
With this in mind, we incorporated continual learning to enhance the performance of our algorithm.
% Training a RL agent to learn the optimal policy from scratch each time is inefficient so we use continual learning to allow the agent to store information from previously solved circuits to reuse in future constructions. 
%
% In this paper we introduce the algorithm used, based on the principles of Probabilistic Policy Reuse (PPR), and how we fitted it to a deep Q-network (DQN). 
%
In this paper, we present the \emph{Probabilistic Policy Reuse with deep Q-learning} (PPR-DQL) framework to tackle this circuit design challenge.
%
% To evaluate the effectiveness of the algorithm, we solve for the circuit solution to reach the Bell State and compare the performances in using the PPR-algorithm or a simple DQN. 
By conducting numerical simulations over various noise patterns, we demonstrate that the RL agent with PPR was able to find the quantum gate sequence to generate the two-qubit Bell state faster than the agent that was trained from scratch. The proposed framework is general and can be applied to other quantum gate synthesis or control problems-- including the 
%efficient, 
automatic calibration of quantum devices.

% The proposed framework implies that the PPR-algorithm successfully increases the efficiency of the design process while simultaneously retaining the properties of circuit optimization ingrained from the previous DRL framework.

% Motivation

% Challenge
%Device noise is constantly changing, training-from-scratch is inefficient. build a model to automatically learn features efficiently (not computationally expensive) 
% Approach
% mention continual learning
% Results & Impact
% Point out major contributions in a few sentences and how the model/framework can benefit QC
\end{abstract}

\maketitle

%\tableofcontents

\section{\label{sec:Indroduction}Introduction}
% \YC{1. Motivate with the recent advances in QC. 2. Certain limitations: drifting noises. (cite) ... 3. Cast into sequential decision making problem --> RL. Need a good control scheme: RL come to the rescue, and mention RL is a good candidate based on recent applications of RL on many control or scientific problems [Hard problems can be solved with RL such as (give examples)] 4. RL can be used in quantum computing problem, mention several relevant works (not only my previous results) -- motivate [5]  5. However, (mention some limitations of these existing works), pointing out the need of a RL for QC framework which is more efficient -- need to retrain the whole model --> inefficient. 6. Describe the major idea of this paper, list major contributions (consider using bullet points) 7. This paper is organized as follows....(just like other papers, refer to my previous paper.)}

% \textbf{Motivation with the QC ideas}\\
Since the conception of quantum computing (QC)~\cite{feynman1985quantum}, many researchers and engineers have sought to use it to solve problems which cannot be practically solved by a classical computer. Theoretically, quantum computing has promised exponential or quadratic speedups for several difficult computational problems that are otherwise intractable on a classical computer \cite{harrow2017quantum, arute2019quantum, nielsen2002quantum}, such as: factorizing large integers \cite{shor1999polynomial} and unstructured database searches \cite{grover1997quantum}. In practice, the applications of QC are largely limited by the capabilities of existing noisy quantum devices \cite{preskill2018quantum} which do not have fidelity-preserving measures such as quantum error correction \cite{nielsen2002quantum, gottesman1997stabilizer, gottesman1998theory} and thus cannot carry out sophisticated quantum computational tasks faithfully. 
Variational quantum algorithms (VQA) have been one approach to dealing with these Noisy Intermediate Scale Quantum (NISQ) devices. 
Recent developments in VQA provide a framework to design near-term quantum applications in various scenarios \cite{cerezo2020variational, bharti2021noisy}. VQA-based applications include solving quantum chemistry problems \cite{peruzzo2014variational} as well as machine learning (ML) tasks such as: function approximation \cite{chen2020quantum, mitarai2018quantum,paine2021quantum,kyriienko2021solving}, classification \cite{mitarai2018quantum,schuld2018circuit,havlivcek2019supervised,Farhi2018ClassificationProcessors,benedetti2019parameterized,mari2019transfer, abohashima2020classification, easom2020towards, sarma2019quantum, stein2020hybrid,chen2020hybrid,chen2020qcnn,wu2020application,stein2021quclassi,chen2021hybrid,jaderberg2021quantum,mattern2021variational,hur2021quantum,li2021recent}, generative modeling \cite{dallaire2018quantum,stein2020qugan, zoufal2019quantum, situ2018quantum,nakaji2020quantum}, deep reinforcement learning \cite{chen19,lockwood2020reinforcement,jerbi2019quantum,Chih-ChiehCHEN2020,wu2020quantum,skolik2021quantum,jerbi2021variational,kwak2021introduction,nagy2021photonic,chen2021variational}, sequence modeling \cite{chen2020quantum, bausch2020recurrent, takaki2020learning,abbaszade2021application}, speech recognition \cite{yang2020decentralizing}, metric and embedding learning \cite{lloyd2020quantum, nghiem2020unified}, transfer learning \cite{mari2019transfer} and federated learning \cite{chen2021federated,yang2020decentralizing,chehimi2021quantum}.
% \textbf{Put a lot of recent application of VQA here}
Despite these achievements, it is still a non-trivial task to design a good quantum circuit architecture for specific computational tasks and such difficulties may hinder a wider acceptance of QC. 

%\textbf{Motivation with the ML ideas, state-of-the-art AI/ML achievements and physical science applications}\\

Concurrently, many advancements have been made in applying classical computing techniques to solve complex tasks. The most prominent example is the significant progress in machine learning (ML) and artificial intelligence (AI) methodologies. Modern deep neural network (DNN)-based models \cite{lecun2015deep,goodfellow2016deep}, which utilize the principles of deep learning (DL), have been applied to a diverse set of areas and have been shown to be highly successful in: computer vision \cite{voulodimos2018deep,simonyan2014very,lecun1998gradient}, natural language processing \cite{bahdanau2014neural,vaswani2017attention,young2018recent,otter2020survey}, and playing the game of Go or many video games with superhuman performance \cite{mnih2015human, silver2017mastering, silver2016mastering, schrittwieser2019mastering, badia2020agent57}, just to name a few. Among these, deep reinforcement learning (DRL) is of our particular interest, as it is designed to solve highly sophisticated sequential decision making tasks \cite{sutton2018reinforcement, bertsekas2021rollout}.

% \textbf{Mention classic AI/ML achievements here}\\

% \textbf{Mention classic RL NAS here}\\

% \textbf{Connect classical NAS and QAS}\\

With powerful RL techniques in hand, it is natural to explore the potential application of RL to address several long-standing problems in quantum computing-- especially the ones which can be reformulated into a sequential decision problem. Recent research has employed the extraordinary capabilities of RL to tackle several challenges in quantum computing such as quantum error correction (QEC) \cite{nautrup2019optimizing,andreasson2019quantum,colomer2020reinforcement,sweke2020reinforcement} and quantum control \cite{niu2019universal,an2019deep,bukov2018reinforcement,fosel2018reinforcement,zhang2019does,palittapongarnpim2017learning,xu2019generalizable,brown2021reinforcement,baum2021experimental}. The common feature among these QC applications is that these tasks can be framed as sequential decision problems. For example, in QEC the goal is to design gate sequences to either encode or decode the quantum states. In the context of the quantum control problem, specific hardware operation sequences (e.g. pulse sequences) are generated to synthesize a desired quantum state or a quantum gate functionality. Another QC problem which can be thought of in this manner is to find a quantum gate sequence in order to achieve a certain computing task. This is called a quantum architecture search problem and is the task we are focusing on in this paper.

% \textbf{Compare q control, ecc and QAS}\\
The goal of a \emph{quantum architecture search} (QAS) algorithm is to find a proper quantum circuit or gate sequence which can be used to achieve a specific quantum computational task. Ideally, the generated circuit architecture should use as few operations as possible to minimize the effect of device noises. Additionally, the QAS algorithm should be hardware-aware, meaning that the algorithm should be able to provide circuit architectures tailored for the specific quantum hardware available. An important distinction is that the QAS problem is different from the quantum control problem \cite{baum2021experimental}, since the latter focuses on generating control pulses on the hardware level \cite{baum2021experimental}. We are interested in QAS algorithms since they can be used in the workflow of quantum compilation, a procedure to transfer quantum algorithms into a gate sequence suitable for a particular quantum device.

The QAS problem has taken inspiration from classical ML techniques and challenges. In classical ML, especially in deep learning, although the models can perform certain tasks pretty well, they are also extremely complex \cite{he2016deep} and contain many components which require manual tuning. Numerous efforts have been made to build automatic procedures to generate well-performing deep neural network architectures for given tasks \cite{zoph2016neural, baker2016designing, cai2017efficient, zoph2018learning, zhong2018practical, schrimpf2017flexible, pham2018efficient, cai2018path, elsken2019neural}, which is known as a neural architecture search (NAS). A common method for NAS is to use RL to sequentially generate or place neural network blocks \cite{zoph2016neural, baker2016designing}. It has been shown that various RL-based methods can generate neural architectures that can achieve performance superior than manually-crafted ones \cite{zoph2016neural, baker2016designing}.
Such achievements led to the exploration of RL-based methods in QAS and recently, researchers have started to apply RL for certain kinds of QAS tasks \cite{ostaszewski2021reinforcement, fosel2021quantum, pirhooshyaran2020quantum, bolens2021reinforcement}.

% YC: the following paragraph can be reused later
% With machine learning, the capability of the quantum computer has been taken in a new direction and the next step would be to improve upon the machine learning itself. Reinforcement Learning (RL) is a subcategory under machine learning that is explained farther in Section \ref{sec:RL}. RL is a very useful tool for problems such as (...). It's been shown in numerous papers (...) that RL has the capability to solve problems that traditional methods could not. With this powerful tool, it was only natural to try to implement it with the quantum computer and there is research which does this (...). 

% \textbf{Mention traditional RL issues}\\
% \YC{Clean this paragraph later}
%\YC{Esther, please read and correct my grammar here.}
Despite RL being a powerful solution to sequential decision making problems, it still has its drawbacks-- a significant issue that the RL framework faces is the need to retrain the agent whenever the environment changes. In the context of hardware-aware QAS, the drifting device noise patterns may reduce the effectiveness of RL-based results. This re-acquisition of information is extremely inefficient due to its redundancy and is precisely the problem we target in this paper.

% \textbf{Proposed our idea (policy reuse)}\\
In this paper, we implement a continual learning scheme to allow the RL agent to learn how to generate quantum circuit architectures in fewer learning episodes when placed in different environments of varying noise-levels. To account for the inefficiency of retraining the agent when the noise pattern is different, we implement an additional aspect of continual learning to our scheme such that the agent will be able to store and reuse the information or policies it learned from previous training environments. 
Concretely, the learned policies will form a \emph{policy library} for future use. When the agent encounters a new environment, the past policies from the library will serve as the \emph{a priori} conditions for the exploration. 
% \YC{Is exploration actually good wording in this context?} 
Instead of exploring the completely random actions, the past policies can be a reasonable foundation for building new policies for an unseen environment. We demonstrate via numerical simulations that such intuitive construction, though very simple conceptually, can largely boost the training efficiency of these quantum architecture searching RL agents.  
%
%\YC{Need to prepare a Fig-1 here, similar to the previous paper but include the continual learning idea..}

% \textbf{List major contributions}\\
Our contributions are the following:
\begin{itemize}
    \item We provide a framework to study how continual reinforcement learning can be applied to quantum architecture search problems. 
    \item We demonstrate building a quantum circuit step-by-step via continual reinforcement learning methods, which allows for new policies to be formed from policies acquired in previous training environments with varying noise levels.
\end{itemize}

% \textbf{paper organization}\\
The paper is organized as follows. 
In \sectionautorefname{\ref{sec:QAS}} we introduce the quantum architecture search (QAS) problem that our agent will solve. 
In \sectionautorefname{\ref{sec:RL}}, we introduce the RL background knowledge used in this work.  
In \sectionautorefname{\ref{sec:CL}}, we introduce the idea of continual learning which will be used to increase the capabilities of existing DQN when dealing with changing environments. 
In \sectionautorefname{\ref{sec:Policy Reuse Deep Q-Learning for QAS}}, we describe the proposed continual DQN used for the QAS. 
In \sectionautorefname{\ref{sec:ExpAndResults}}, we describe the details of our experimental methodology and our hyperparameter settings.
In \sectionautorefname{\ref{sec:Results}}, we present the numerical simulation results of the continual DQN as well as the results of the DQN without continual learning for comparison. 
Finally we discuss the results in \sectionautorefname{\ref{sec:Discussion}} and offer concluding remarks in \sectionautorefname{\ref{sec:Conclusion}}.

\section{\label{sec:QAS}Quantum Architecture Search (QAS) Problem}
%
% \textbf{Intro to the QAS problem setting}\\
%
\emph{Quantum architecture search} (QAS) is an algorithmic procedure designed to find a series of quantum operations or quantum gates to achieve a predefined target such as generating the desired quantum state from an initial state.
Concretely, given an initial state $\ket{0 \cdots 0}$ and the target state $\ket{\Psi}$, the goal is to find out a proper quantum gate sequence such that after the operation of these gates, the initial state can be transformed into the target state within a certain error tolerance. 
%
% \textbf{Describe the high-level RL procedure here}\\
%
We will be using RL to achieve this goal. In the context of RL training, the environment $\mathcal{E}$ is the quantum computer or quantum simulator. In this work, we chose to use a quantum simulator (the details of which can be found in \sectionautorefname{\ref{sec:ExpAndResults}}). In using the quantum simulator, we can define various noise or error pattern to evaluate and test the capability of our proposed continual RL framework. Another reason for using a quantum simulator over a real quantum computer is that it is currently impractical to directly train a large number of episode through cloud-based quantum platforms.
% \textbf{Describe the high-level RL procedure and observation, fidelity etc.... ref to the old paper}
%

The observations, used by the RL agent to process the state of the system, are composed of several Pauli measurements (e.g. Pauli-$X$, $Y$ and $Z$ expectation values from each qubit) and the agent calculates a reward value from the \emph{fidelity} (similarity) of the generated state to the target state. 
In each time-step, the RL agent chooses an action $a$ from the action set $\mathcal{A}$, which includes various quantum gates (one- and two-qubit gates) and the index of the qubit they will operate on. 
%
%\textbf{\sout{Need paraphrasing the following sentences!} -- done}
The agent updates the quantum circuit with the selected action and then the environment $\mathcal{E}$ executes the new circuit and calculates the \emph{fidelity} of the current state to the target state. 
%
%\textbf{\sout{Need paraphrasing the following sentences!} -- done}
%\YC{I modified this sentence a little bit. Please check.}
If the fidelity is greater than a predefined value-- meaning that the quantum state is within an accepted error tolerance to the target state-- then the episode ends there and a large positive reward is fed back to the agent. 
If, however, the fidelity is below the predefined value, the RL agent will be penalized with a small negative reward before moving onto the next time-step.
During each time-step, the environment $\mathcal{E}$ returns Pauli expectation values of each qubit as the observations (or states), to the RL agent. For an $n$-qubit system, the dimension of the state or observation vector is subsequently $3n$. 
%\EY{as there are 3 Pauli metrics?}.
%
The procedure continues until the RL agent reaches either a sufficiently high value of fidelity or the maximum allowed steps. 
%
% The deep Q learning with probabilistic policy reuse is used to optimize the RL agent.
Various algorithmic procedures such as the policy gradient method or deep Q-learning can be used to optimize the RL agent.
%
% \textbf{Discuss our previous work and limitations} \EY{added one sentence for the limitations}

For example, the previous work \cite{kuo2021quantum} has demonstrated that DRL algorithms were able to sample from a set of basic quantum gates in order to incrementally construct a quantum gate sequence that yielded a desired output state. This capability was seen in quantum circuit architectures consisting of two- and three-qubits and the target state of that paper was the Bell state (two-qubit) and the GHZ state (three-qubit). Conceptually, the Bell state and GHZ state are equivalent states for systems with different numbers of qubits. This state was picked due to it's prominent usage in quantum computing and quantum communication for the purpose of quantum entanglement. 

Although the DRL algorithms in the previous paper \cite{kuo2021quantum} were successful, there is critical limitation to that framework. The requirement to retrain the RL agent for each new task (such as environments of varying noise levels) reduces the usefulness of the method in quantum device calibration. %\YC{Esther, please review the above sentence.}
In order to address this obstacle, we incorporated a continual learning framework (described in the \sectionautorefname{\ref{sec:CL}} and \sectionautorefname{\ref{sec:Policy Reuse Deep Q-Learning for QAS}}) to the DRL training procedure such that the RL agent can leverage previously gained knowledge when encountering a new environment. 
% In the (\sectionautorefname{\ref{sec:CL}}), we will go over the key principles of the concept as well as take a closer look at the role it will serve in this project.

%
In this paper, we retained the Bell state as our target state, and adopted a similar assessment (reward) scheme as before \cite{kuo2021quantum} to now evaluate the continual DRL algorithm which uses probabilistic policy reuse. 
The Bell state has the following circuit solution for a two-qubit system:
\newline
\begin{figure}[!htbp]
\begin{center}
\begin{minipage}{10cm}
\Qcircuit @C=1em @R=1em {
\lstick{\ket{0}} & \gate{H}  & \qw        & \ctrl{1}       & \qw      & \qw \\
\lstick{\ket{0}} & \qw       & \qw        & \targ          & \qw      & \qw
}
\end{minipage}
\end{center}
\caption[Quantum Circuit for Bell state.]{{\bfseries Quantum circuit for the Bell state.}
}
\label{Fig:circuitForBellState}
\end{figure}
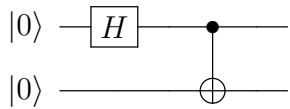
\newline \tab \tab
% The use of the DRL algorithms is accompanied by a means of assessment, as mentioned earlier, which is a scoring system. We chose to keep the same scoring metric as the previous paper as it proved to be a good marker of the performance of the algorithms and this score system will be explained further in the DRL subsection below. 
% The reward scheme used in this work follows the previous paper \cite{kuo2021quantum} and will be described later in  \equationautorefname{\ref{eqn:score_calculation}}.

\figureautorefname{\ref{Fig:circuitForBellState}} shows the optimal circuit solution which the RL agent will be trying to construct incrementally in environments with different noise-conditions.

\section{\label{sec:RL}Reinforcement Learning}
%\YC{Should introduce RL first, then say that the RL which uses deep neural networks are called DRL}
%\YC{Describe the deep Q learning in this section}
%\YC{\textcolor{red}{the following RL description should be paraphrased to avoid self-plagiarism}}
%\EY{Incorporate DRL from pages 8-9 to differentiate against plagiarism}
%

\emph{Reinforcement learning} (RL) is a method of machine learning which allows the \emph{agent} to learn by optimizing a score based on its interactions with the environment ~\cite{sutton2018reinforcement}. The \emph{agent} interacts with an \emph{environment} $\mathcal{E}$ over a number of discrete time steps. At each time step $t$, the agent receives a \emph{observation} or \emph{state} $s_t$ from the environment $\mathcal{E}$ and feeds that information into its current \emph{policy} $\pi$ in order to choose an \emph{action} $a_t$ from a set of possible actions $\mathcal{A}$. The policy $\pi$ is a function which maps the state or observation $s_t$ to an action $a_t$. In general, the policy can be stochastic-- for a given state $s$, the action output can be a probability distribution $\pi(a_t|s_t)$ conditioned on $s_t$. After executing the action $a_t$, the agent receives the state of the next time step $s_{t+1}$ and based on that state it gets a scalar \emph{reward} $r_t$. This reward is what affects the score of the agent for this task. It can either be a positive quantity (which indicates that a good move was made) or a negative quantity (which can be thought of as a penalty/deduction for a bad choice). This process is continued until the agent reaches the terminal state or a pre-defined stopping criteria (e.g. the maximum steps allowed). An \emph{episode} is defined as an agent starting from a randomly selected initial state and following the aforementioned process either all the way through until it gets to the terminal state or until it reaches the stopping criteria.

We define the total discounted return from time step $t$ as $R_t = \sum_{t'=t}^{T} \gamma^{t'-t} r_{t'}$, where $\gamma$ is a discount factor in $(0,1]$. The role of $\gamma$ is have a parameter to control how future rewards are weighted to the decision making function. When a large $\gamma$ is considered, the agent weighs the future reward more heavily. On the other hand, with a small $\gamma$, future rewards are quickly ignored and immediate reward will be weighted more. The value of $\gamma$ can vary across time steps as well and this is what was done in this paper (\sectionautorefname{\ref{sec:ExpAndResults}}). 
The goal of the agent is to maximize the expected return from each state $s_t$. The \emph{action-value function} or \emph{Q-value function} $Q^\pi (s,a) = \mathbb{E}[R_t|s_t = s, a]$ is the expected return for selecting an action $a$ in state $s$ based on policy $\pi$. The optimal action value function $Q^*(s,a) = \max_{\pi} Q^\pi(s,a)$ will give the maximal action-value across all possible policies. The agent's expected return from following policy $\pi$ at state $s$ is given by $V^\pi(s) = \mathbb{E}\left[R_t|s_t = s\right]$, which is called a value function. There are various RL algorithms which are designed to find the policy that maximizes the value function. This framework of using RL algorithms to maximize the value function is called \emph{value-based} RL. 

%The area where RL falls short is its limitation on the environment. RL algorithms struggle with environments that are not well-defined/observed prior to running the algorithm or environments that have a great number of observable input states \cite{mnih2015human}. That is why we are motivated to combine deep learning into the scheme (\sectionautorefname{\ref{sec:RL} B}) so that the agent can take into account previous moves or actions it has made in the past. 

%\emph{Deep reinforcement learning} (DRL) involves using deep neural networks to conduct reinforcement learning. In this paper, we utilize it to implement deep Q-learning (see section \ref{sec:DQL}), which has been shown to be successful in . 
%\EY{Motivate this method?}
%Furthermore, DRL has 
%\EY{reference some examples here: why we use Neural network as function approximator --> much more scalable -- still works with a large number of qubits}

%
\subsection{Q-Learning}
%\YC{\textcolor{red}{the following Q-learning description should be paraphrased to avoid self-plagiarism}}
%
Q-learning \cite{sutton2018reinforcement} is a model-free approach to the RL algorithm, meaning it does not attempt to learn about the underlying transition dynamics. Before the learning process begins, the value function $Q$ is randomly initialized. Then, at each time step $t$, the agent selects an action $a_t$ (using a selection function such as an $\epsilon$-greedy policy derived from $Q$). The agent then observes the reward $r_t$ that resulted from this action and enters a new state $s_{t+1}$. After the agent enters the new state $s_{t+1}$, $Q$ is updated with the learning rate $\alpha$. This makes Q-learning an \emph{off-policy} learning technique since it updates its Q-values using the observed reward $r_t$ and the \textcolor{black}{maximum reward $\max _{a} Q\left(s_{t+1}, a\right)$ of the next state $s_{t+1}$ over all possible actions $a$.} The updating is done according to the following benchmark formula:

\begin{align}
  Q\left(s_{t}, a_{t}\right) \leftarrow  Q\left(s_{t}, a_{t}\right)
  +\alpha\left[r_{t}+\gamma \max _{a} Q\left(s_{t+1}, a\right)-Q\left(s_{t}, a_{t}\right)\right].
\end{align}
\subsection{\label{sec:DQL}Deep Q-Learning}
%\YC{\textcolor{red}{the following deep Q-learning description should be paraphrased to avoid self-plagiarism}}
%
 The action-value function $Q(s, a)$ can be represented by a two-dimensional table with $s \times a$ total entries, that is, the number of possible states times the number of possible actions. However, when the state space or the action space $\mathcal{A}$ is large or even continuous, the tabular method is unfeasible. In these situations, the action-value function can be represented with function approximators such as neural networks~\cite{mnih2016asynchronous, mnih2015human}. This neural-networks-based reinforcement learning is called \emph{deep reinforcement learning} (DRL).

The usage of neural networks as function approximators in order to represent the $Q$-value function has been studied extensively~\cite{mnih2016asynchronous, mnih2015human} and has been shown to be successful in many tasks such as playing video games. In this setting, the action-value function $Q(s, a;\theta)$ is parameterized by $\theta$, which can be derived using a series of iterations from a variety of optimization methods adopted from other machine learning tasks. The simplest form of DRL is $\emph{deep Q-learning}$. For deep Q-learning, the goal is to directly approximate the optimal action-value function $Q^*(s, a)$ by minimizing the mean square error (MSE) loss function: 
\begin{equation}
L(\theta) = \mathbb{E}[(r_t + \gamma  \max_{a'} Q(s_{t+1},a';\theta^-) - Q(s_t,a_t;\theta))^2].
\end{equation}

Here, the prediction is $Q(s_t,a_t;\theta)$,
where $\theta$ is the parameter of a policy network (the network from which we extract our solution)
and the target is $r_t + \gamma  \max_{a'} Q(s_{t+1},a';\theta^-)$, where $\theta^-$ is the parameter of the target network (created to help stabilize the optimization procedure) and 
$s_{t+1}$ is the state encountered after performing action $a_t$ at state $s_t$.

In DRL, it is ideal for the loss function to converge but oftentimes it is very difficult to get it to do so. In fact, the loss function will usually be divergent when a nonlinear approximator like a neural network is used to represent the action-value function~\cite{mnih2015human}. There are several possible reasons for why this happens. When the states or observations are serially correlated along the temporal trajectory --thereby violating the assumption that the samples are independent and identically distributed (IID)-- the $Q$ function varies dramatically and subesquently changes the policy at a relatively large scale. Besides the states being correlated, another reason for divergence could be due to having a large correlation between the action-value $Q$ and the target values $r_t + \gamma  \max_{a'} Q(s_{t+1},a')$. Unlike in supervised learning where the targets are given and invariant, the settings of DRL
% deep reinforcement learning
allow targets to vary with $Q(s,a)$, causing $Q(s,a)$ to chase a nonstationary target.

The \emph{deep Q-learning} (DQL) or \emph{deep Q-network} (DQN) presented in the work \cite{mnih2015human} addressed these issues by implementing two mechanisms:
\begin{itemize}
\item \emph{Experience replay}: To perform experience replay, one stores each transition the agent encounters a tuple in the following form:
$(s_t, a_t, r_t, s_{t+1})$ for each time step $t$. To update the $Q$-learning parameters,
one randomly samples a batch of experiences from the replay memory and then performs gradient descent with the following MSE loss function:
$L(\theta) = \E[(r_t + \gamma  \max_{a'} Q(s_{t+1},a';\theta^-) - Q(s_t,a_t;\theta))^2]$, where the loss function is calculated over the previously sampled batch from the replay memory. The key feature of adding the experience replay is to lower the correlation of the inputs for training the $Q$-function.
\item \emph{Target Network}: Here $\theta^-$ is the parameter of the target network and these parameters are only updated after a certain number of finite time steps. This setting helps to stabilize the $Q$-value function training since the target is relatively stationary compared to the action-value function.
\end{itemize}

In the next section, we will introduce the proposed continual DQN which is capable of reusing previous policies. The major components of DQN such as experience replay and target network will be preserved.

%\EY{More citations and discussion of why this was chosen}
%\EY{Make it simple and include mathematical notations to be carried through the paper}
%DRL is a member of the general family of machine learning. Machine learning involves having a computer learn to solve problems rather than inputting the solutions into the code directly. 

%To summarize, DRL is a fairly new technique \cite{mnih2015human} that bridges the fields of deep learning and reinforcement learning (RL) to be more powerful than either alone. The RL component uses a score in order to assess the performance of the learning agent. The scoring system is comprised of rewards (positive quantities added to the score which indicate a good move) or penalties/deductions (negative quantities that are taken from the score when there’s a bad performance). The area where RL falls short is its limitation on the environment. RL algorithms struggle with environments that are not well-defined/observed prior to running the algorithm or have a great number of observable inputs states \cite{mnih2015human}. To counter this, deep learning is combined into the scheme to take into account previous moves or actions the agent has already made. 

With DRL, the agent is encouraged to move towards the optimal solution over time while considering previous actions. The DRL agent associates its performance with the score it receives every learning episode and aims to improve upon that score as episodes progress.
From the system that was carried over from the previous paper \cite{kuo2021quantum}, the scores in these experiments are calculated by taking the quantum state fidelity (in other words, how close the observed output state was to our target Bell state) and subtracting the penalty term (0.01) for each step taken. This can be seen in the following equation (\equationautorefname{\ref{eqn:score_calculation}}): 
\begin{equation}
\label{eqn:score_calculation}
Episode\ Score = State\ Fidelity - (0.01) \cdot (number\ of\ steps\ taken)
\end{equation}
Having a higher episode score meant that the quantum state is close to the Bell state (high state fidelity) and that not many gates were used in the circuit construction (small number of steps).

\section{\label{sec:CL}Continual Learning}
%\YC{Intro the idea of CL}
%
% \EY{Source}
Recent advances in deep RL have reached several extraordinary achievements through beating their best human competitors on a variety of challenging tasks \cite{mnih2015human,silver2016mastering,silver2017mastering}. However, within existing approaches, a computational agent is trained to master a narrow task of interest. 
% Such approaches largely limit the generalizability of deep RL agents.
%
Moreover, untrained computational agents require significantly more training episodes to reach a good performance compared to their human competitors. After training, it is hard to be generalized to other similar tasks, even simple RL ones \cite{bengio2020interference}. 
In contrast, humans have the capability to continually learn new skills and incrementally adapt to unseen scenarios over the duration of their lifetime. Such ability is called continual learning.
\emph{Continual learning} (CL) indicates the learning process which is the constant and incremental development of increasingly complex behaviors. This includes the process of developing more and more sophisticated behaviors on top of skills or knowledge already learned \cite{ring1998child}. CL is closely related to topics such as \emph{lifelong learning}, \emph{online learning} and \emph{never-ending learning}.
%
% \YC{Should merge with the following texts}

%
%\EY{Go back to this section too}
%\EY{More citations and discussion of why this was chosen}
%\EY{List recent advances to emphasize what CL is}
%\YC{I think CL is not something newer than RL, it is a framework to help extending the RL capabilities?}
%Continual learning is even newer than DRL and has largely only existed in the realm of theory but was seen as the natural progression after RL \cite{khetarpal2020towards}. 
The principle of continual learning centers around the idea that the agent does not stop learning information as more tasks are given to it. In non-continual learning, the agent is given a separate training sequence to run in order to learn information that will be applied to future tasks. However with continual learning, the information gained from training, or \emph{pre-gained knowledge}, is being expanded with each task that the agent completes. After the agent completes a new task, any information gained from running it will be added to the knowledge it has already gained during the training run. This means that the knowledge that the agent already has going into each new task, will be expanding indefinitely as more and more tasks are performed. 

%\YC{motivate the use of CL. e.g. constantly changing noise patterns?}

This was a crucial capability which motivated us to implement CL for the purposes of adapting the DRL framework to function in differing environments, specifically environments with varying noise patterns. Implementing CL for QAS allows the RL agent to recall previous circuit designs that it has generated and reuse them as it sees fit. This not only expedites the design of circuits targeting similar states, but also the creation of the correct circuit under different environmental conditions (e.g. noise-levels).

\section{\label{sec:Policy Reuse Deep Q-Learning for QAS}Probabilistic Policy Reuse with Deep Q-Learning for QAS}
%\EY{include references to the appendix}
%\EY{fix the later part of this section}
%\YC{Describe the basic elements of our framework}
%
To build an efficient RL framework for quantum architecture search problems, we contribute the \emph{Probabilistic Policy Reuse with deep Q-learning} (PPR-DQL) algorithm-- to be referred to simply as the PPR-algorithm in future sections. Our method is inspired by Fernandez's work \cite{fernandez2006probabilistic} which considers the implementation of the policy reuse framework using classic tabular Q-learning. 
In this paper, we extend upon the original concept to include some of the more modern developments in deep Q-learning. Specifically, we incorporate the following two crucial components of deep Q-learning into the policy reuse framework: \emph{experience replay} and \emph{target network}. In this method, previously learned policies are used as probabilistic biases when the learning agent considers the following three choices: 1) the exploitation of the ongoing learned policy, 2) the exploration of random actions and 3) the exploitation of previously learned policies.
%
%\EY{With the new ordering of this section, check the flow of information and make sure that any part which mentions the experiments is not included here (that comes afterwards)}
%\EY{Introduce framework and how it fits the problem we're looking at. For the ideas, the specifics of the hyperparameters and number of cases will go in experimental section}
%
The PPR is implemented outside of the DQN, using its capabilities and structures such as the experience replay and the target network. Training the agent from scratch can be thought of as utilizing the PPR-algorithm without any pre-loaded policies. That is to say, the training-from-scratch simulations also use a DQN to learn and select the appropriate action. With the PPR-algorithm, the difference is that there are now pre-loaded models available to be considered, which are also neural networks, in each run of the algorithm. 
%
%\newline \tab
%\tab In order to implement the PPR, two main components are added to the pre-existing code \YC{In general, we will not talk about code implementation details here...}. First, a policy library is constructed to store the policies currently available for consideration. The library object itself is a python list that contains references to the DQNs of the previous models (i.e. the appropriate weights and biases saved from those models). 

In order to implement the PPR, two main components are created. The first is a policy library $L$ which holds the policies. Along with the policy library, the second component we create is an algorithmic procedure to decide whether to use a previously solved policy stored in the library or have the agent explore on its own. The exploration option is preferable when the new task given to the algorithm is vastly different from any of the previously solved models (a case where reusing an old policy may deter the algorithm more than it would help it to arrive at the right circuit). In the appendix (\appendixautorefname{\ref{sec:Appendix}}) are psuedocodes that further detail the inner workings of the algorithm. 
% \EY{How to reference appendix?}. 
%\EY{In-line equation instead of box, write Probability vector P is calculated according to the softmax function with temperature parameter tau }

To select a previous policy from the policy library to examine, we use the softmax function (\equationautorefname{\ref{eqn:softmax}}). This function takes the current rewards vector $W$ and a temperature parameter $\tau$ to return a probability vector $P$. This probability vector is what is used to select a policy from the policy library to look at in the PPR-algorithm (\algorithmautorefname{\ref{policy_reuse_alg}}). 
% \EY{Algorithm 5 cite}.

\begin{equation}
    P(\Pi_{j}) = \frac{e^{\tau W_{j}}}{\sum_{p = 0}^{n}e^{\tau W_{p}}}
    \label{eqn:softmax}
\end{equation}

The q-learning algorithm (\algorithmautorefname{\ref{q_learn_alg}}) works in conjunction with the $\pi$-exploration algorithm (\algorithmautorefname{\ref{policy_reuse_pi_alg}}). The q-learning algorithm is used in the PPR-algorithm when the softmax function decides it is best to perform a greedy action selection from the new policy $L_{0}$. 
%assigns what is currently stored as the new policy to be the 'previous' policy under consideration. This is how the algorithm chooses against following one of the actual previous policies and is the favorable option in the scenario mentioned earlier where the previous models are not very beneficial to the solving of the current task. 
The $\pi$-exploration algorithm in \algorithmautorefname{\ref{policy_reuse_pi_alg}}, on the other hand, takes into account a past policy that was loaded into the library (again, the one chosen depends on the softmax function). It samples a probability which it compares against the hyperparameter $\psi$ to decide whether it wants to select its action using the past policy chosen or the current new policy $L_{0}$ being developed during this experiment.
One attribute that the $\pi$-exploration and q-learning algorithms share is the sampling of transition pairs from the replay memory $\mathcal{D}$. These transition pairs are then used to optimize $L_{0}$, thus ensuring that previous knowledge is being used to guide the development of the new policy.
%
%Through the $\pi$-exploration algorithm, transition pairs from both the old policies $L_{i}$ and the new policy $L_{0}$ are stored in the replay memory $\mathcal{D}$. This allows for the sampling of previous transition pairs from $\mathcal{D}$ even when using the q-learning algorithm, thus ensuring that previous knowledge is always being reused. 

Besides the implementation of the PPR-algorithm on a DQN rather than on tabular Q-learning as Fernandez demonstrated \cite{fernandez2006probabilistic}, we also chose to use different action selection policies. After the DQN is traversed and each action is given a certain probability, there are two action policies that are used for different scenarios. The first is the simple greedy policy which tells the agent to choose whichever action has the greatest probability of success. The other method is the epsilon-greedy policy which assigns yet another probability in choosing the action with the highest probability of success. Although Fernandez  \cite{fernandez2006probabilistic} utilized the epsilon-greedy selection policy in their PPR scheme when exploring a new policy, we found that the simple greedy policy converges to the optimal solution faster and chose to use that in its place. However, we kept the epsilon-greedy when performing the experiments from scratch as it out-performs the greedy policy there.
% \newpage
%Strictly using the q-learning algorithm can be thought of as running the experiment from scratch as it never refers to the other policies in the library.

To optimize the neural networks (specifically, the policy network), we used the ADAM optimizer \cite{kingma2014adam}. Further details on the hyperparameters used can be found in \sectionautorefname{\ref{sec:PPRAlg}}.

\section{\label{sec:ExpAndResults}Experimental Methods}
\subsection{Experimental Setup}
In this section we delve into the implementation of the DRL and continual learning components and how we gauged their effectiveness. To test whether incorporating continual learning improved the agent’s performance in finding the correct quantum circuit architecture, we compared test cases where we do not use the PPR-algorithm to ones where we do use it. Specifically, we looked at whether loading in the models that were solved previously in different noise environments allowed the agent to solve for the Bell state circuit  \figureautorefname{\ref{Fig:circuitForBellState}} faster than if it was tasked to solve it from scratch.  

As stated previously, to test the agent’s ability we maintained the same target state for each simulation and observe the performance of our training algorithms under multiple noise environments. The details for how we setup the noise can be found in \sectionautorefname{\ref{sec:Noise}}. As the noise complexity of an environment is increased, it becomes more difficult for the agent to learn from scratch, but it is expected that using the PPR-algorithm improves its capability to solve these harder problems. The general scheme for the experiments is outlined in \tableautorefname{\ref{tab:noiseCombo}}.
%\EY{Take out environment 6 and drop the last and 2nd and 3rd rows of the table}
\begin{table}[!htbp]
\begin{tabular}{|c|c|c|c|c|c|c|}
\hline
Gates & Env. 0 & Env. 1 & Env. 2 & Env. 3 & Env. 4 & Env. 5 \\ \hline
$X$     & -      & 0.01   & 0.01   & 0.01   & 0.005  & 0.01   \\ \hline
$H$     & -      & -      & 0.01   & -      & 0.005  & 0.01   \\ \hline
$CNOT$  & -      & -      & -      & 0.01   & 0.005  & 0.005  \\ \hline
\end{tabular}
\caption{{\bfseries The sequence of different environments and their respective noise levels.} The table gives the values of the probability of error and which gate it corresponds to for each environment.}
\label{tab:noiseCombo}
\end{table}
%
%\EY{We looked at environment 1 and ran two experiments, one where the conducted training-from-scratch and the other with policy reuse and reference the policies}
%

\begin{figure}[htp]
    \centering
    \scalebox{0.4}{
    \includegraphics{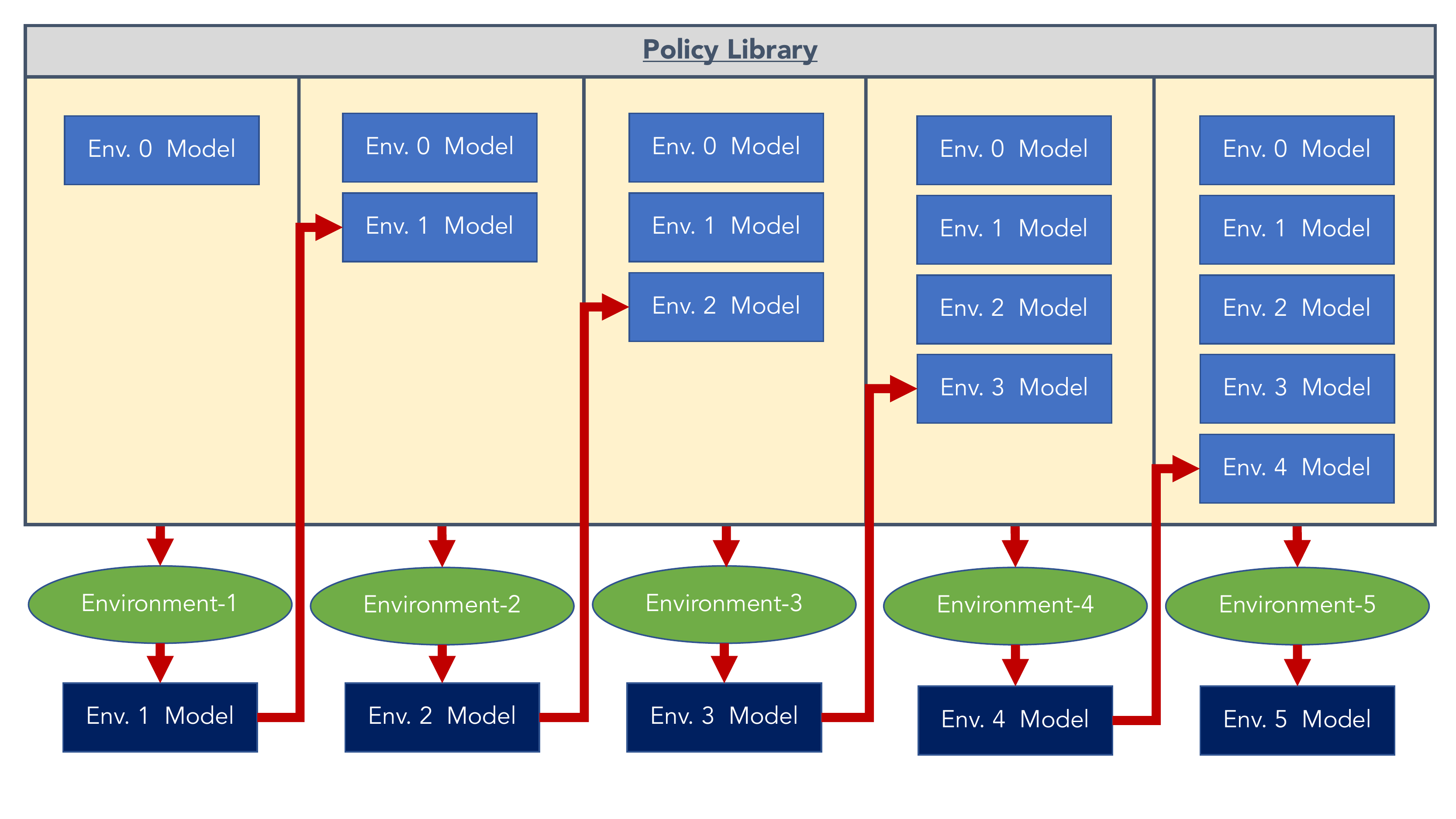}
    }
    \caption{{\bfseries Overview of Policy Reuse in quantum architecture search problems.} This figure depicts the PPR-DQL framework applied to the five environments with varying noise levels. As shown, the policy learned for each environment is fed  into the policy library $L$ to be used for the following environments.}
    \label{Fig:ppr}
\end{figure}

To incorporate the PPR algorithm properly, we first load the model (or policy) solved in \texttt{Environment-0} --which is the one without noise-- to solve for the Bell State in \texttt{Environment-1} --where there is synthetic noise on the $X$ gate. By adding the noise on the $X$ gate, we make it more difficult for the agent to discern the system's true state which makes it a harder environment since it will be harder to solve for the right circuit. 
%
%\YC{Need to confirm the experiment order here. and also the results figure.}
Then after the run for \texttt{Environment-1} is finished, we load the generated policy resulting from it to join the original model trained with the noise free environment, \texttt{Environment-0}, in the policy library in order to solve the next environment, \texttt{Environment-2} --which now has noise on two gates: the $X$ and $H$ gates. 
After that is solved, we repeat the process again of loading the solved model with the other previously solved ones to explore the next environment, \texttt{Environment-3}, and so on. This leads to a framework like the diagram depicted in \figureautorefname{\ref{Fig:ppr}}.

\subsection{\label{sec:Noise}Synthetic Noise}
To vary the environments for each experiment, we introduce a synthetic noise which can be applied to gates of our choosing. There are two types of errors: gate errors and measurement errors. Gate errors are the result of imperfections while performing the quantum operation and measurement errors are ones that occur during the quantum measurement process. The noise we use for the gate error is a depolarizing noise model and it is synthesized through the use of a Qiskit package (Aer). For the gate error, the model switches the state of a designated qubit to a random state with probability $p_{gate}$. For the measurement error, the model assigns a random flip between $0$ and $1$ to be the probability of error $p_{meas}$ right before the measurement is taken . We refer the interested readers to the previous work \cite{kuo2021quantum} for details of the implementation.
%\YC{Do we need to mention the results here? I think we present the results in the next section.}
For our settings, we chose the probability of a measurement error $p_{meas}$ be 0.01 across all the simulations and only varied the probability of the gate error $p_{gate}$ across the environments to change the difficulty. The decision to have a consistent probabilistic measurement error was so that we could isolate the individual gate noise errors as the contributing factor for how well the algorithm performs.

\subsection{\label{sec:DQN}Deep Q-Network Setup}
The overall DQN structure used is the same for all the simulations (encompassing both the PPR-algorithm and the training-from-scratch). For a given learning episode, the agent observes its current state at each time step $t$ and samples from an action set $\mathcal{A}$ of quantum gates to choose one as the action which it believes will drive it towards the correct solution. The action set is comprised of the Pauli-$X$ ($X$), Pauli-$Y$ ($Y$), Pauli-$Z$ ($Z$), Hadamard ($H$), Controlled Not ($CNOT$), and $\frac{\pi}{4}$-Rotation gates. Placing a gate is the equivalent of choosing an action but an important detail to note is that in the initialization of each gate, we also require a corresponding index when working with multiple qubits. The index will represent which qubit the gate should be placed for so the agent can distinguish between actions and apply the right gate to the right qubit. 

% To choose an appropriate action, probabilistic weights are assigned to each possible choice and the agent searches for the best one by going through the DQN. 
The implementation of our DQN is based upon the packages in PyTorch \cite{pytorch}. We designed the DQN in this project to be simple but efficient, opting to use 2 hidden layers and a linear neural network. The input to our DQN is the state of the system $s$ (which is encoded as a vector of the Pauli-$X$, $Y$, and $Z$ observables). The dimension of the input is proportional to the number of qubits $n$, specifically it is $3n$. The dimension of the output is given by the following formula:
\begin{equation}
\mathbb{G}=\bigcup \limits_{i=1}^{n}  \left\{U_i\left(\pi/4\right), X_i, Y_i, Z_i, H_i, CNOT_{i,(i+1)(mod2)}\right\},
\label{eq:action gates}
\end{equation}
where $\mathbb{G}$ is the set of allowed quantum operations, which is our action space. $U(\pi/4)$ is the $\frac{\pi}{4}$-Rotation gate, $X$ is the Pauli-$X$ gate, $Y$ is the Pauli-$Y$ gate, $Z$ is the Pauli-$Z$ gate, $H$ is the Hadamard gate, and $CNOT $ is the Controlled-Not gate.

In this paper we simulated this state $s$ by using a quantum simulator rather than measuring from a real device but our framework is structured such that real measurements can be used as the appropriate technology becomes available. After the state input is passed to the DQN, the output of the DQN holds the values for each of the actions $a$ and the action with highest action-value is selected. 

%traversing the network will result in associating a probability to each of the possible actions $a$ (the gates) and allow the agent to choose the best action to move forward with according to the principles outlined in \sectionautorefname{\ref{sec:RL}}. 

\subsection{\label{sec:gym}Customized OpenAI Gym}

%\EY{This and synthetic noise paragraph: mention that based on previous work we are able to generate the different noise environments needed for this experiment (and cite paper). Condense this paragraph.}
The quantum environments were built by tools developed in the previous paper \cite{kuo2021quantum}. To apply our algorithm to this simulated quantum environment, we created a customized OpenAI Gym environment. With OpenAI gym, we were able to set: any desired target quantum state, a fidelity threshold, and the quantum computing backend for the simulation (which can be a real device or simulator software). Additionally, we were able to customize our noise patterns (detailed in Section \ref{sec:Noise}).

\subsection{\label{sec:PPRAlg}Probabilistic Policy Reuse Algorithm Hyperparameters}
For the PPR-algorithm, we assigned the following values to our hyperparameters: For the temperature parameter $\tau$ (which is updated during each episode), we set the initial value to be $0$. It increases incrementally with the number of episodes that have passed by $\delta\tau$, which we had as $0.01$. We set the number of total episodes $K$ to be $1000$. This value was chosen based on the performance of the algorithms in \cite{kuo2021quantum} and the maximum number of steps in a given episode, $H$, was set to be $20$ due to the inaccuracies that would arise if we tried to put down more than $20$ gates, based on the current state of quantum devices. For the replay memory $\mathcal{D}$ we initialized a capacity of $10,000$ transition pairs and for the ADAM optimizer we used $\beta_{1} = 0.9$ and $\beta_{2} = 0.999$ in this work. The learning rate for the ADAM optimizer $\eta$ was set to $0.001$. Additionally, we have the following hyperparameters for the $\pi$-exploration algorithm: We assign the initial value of $\psi$ (the probability of following a previous policy) to be $1$ and the value for $\nu$ (the decay factor of $\psi$) to be $0.95$.

\section{\label{sec:Results}Results}
\subsection{Environment 0: Noise-Free}
% \newpage
% \advance\vsize by 2cm % Advance page height
% \advance\voffset by 2cm % Shift top margin
\begin{figure}[!htbp]
     \begin{subfigure}[t]{1\textwidth}
     \centering
         \scalebox{0.5}{
            \includegraphics{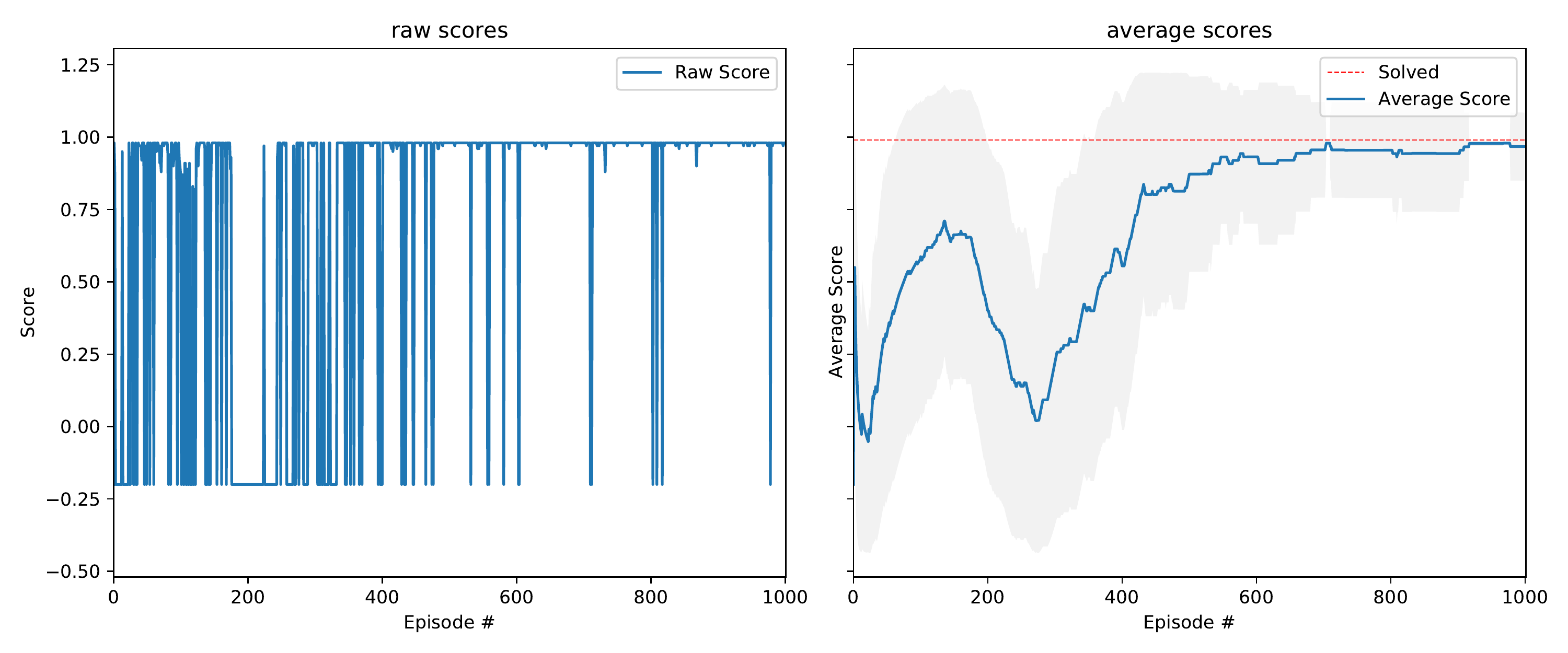}
            }
        
    \end{subfigure}
    \caption{\label{fig:noisefree} {\bfseries  \texttt{Environment-0} (Noise-Free) training-from-scratch simulation}
        }
    \label{fig:env0}
\end{figure}
%\newpage
% \newpage
% \par\vfill\break % Break Last Page
% \advance\vsize by 2cm % Advance page height
% \advance\voffset by 2cm % Shift top margin

To get the first policy for the PPR-algorithm such that its library would not be empty for the next task, we ran a simulation to solve for the Bell State circuit in a noise-free environment. This was done with just the DQN scheme as it is the first policy and there were no previous models to load into the library. From the results in \figureautorefname{\ref{fig:env0}}, we see that the DQN converges to the correct circuit solution around the $500^{th}$ episode when the score returned is close to $0.98$. The score is calculated according to \equationautorefname{\ref{eqn:score_calculation}}. This performance will serve as the benchmark (a control case of no noise) for comparisons to the more difficult environments, and the policy generated after the completion of the $1000^{th}$ episode will be added to the policy library to train the next policies.

\subsection{Environment 1: $0.01$ Noise on the $X$ Gate}
\begin{figure}[!hbp]
     \begin{subfigure}[t]{1\textwidth}
     \centering
         \scalebox{0.5}{
            \includegraphics{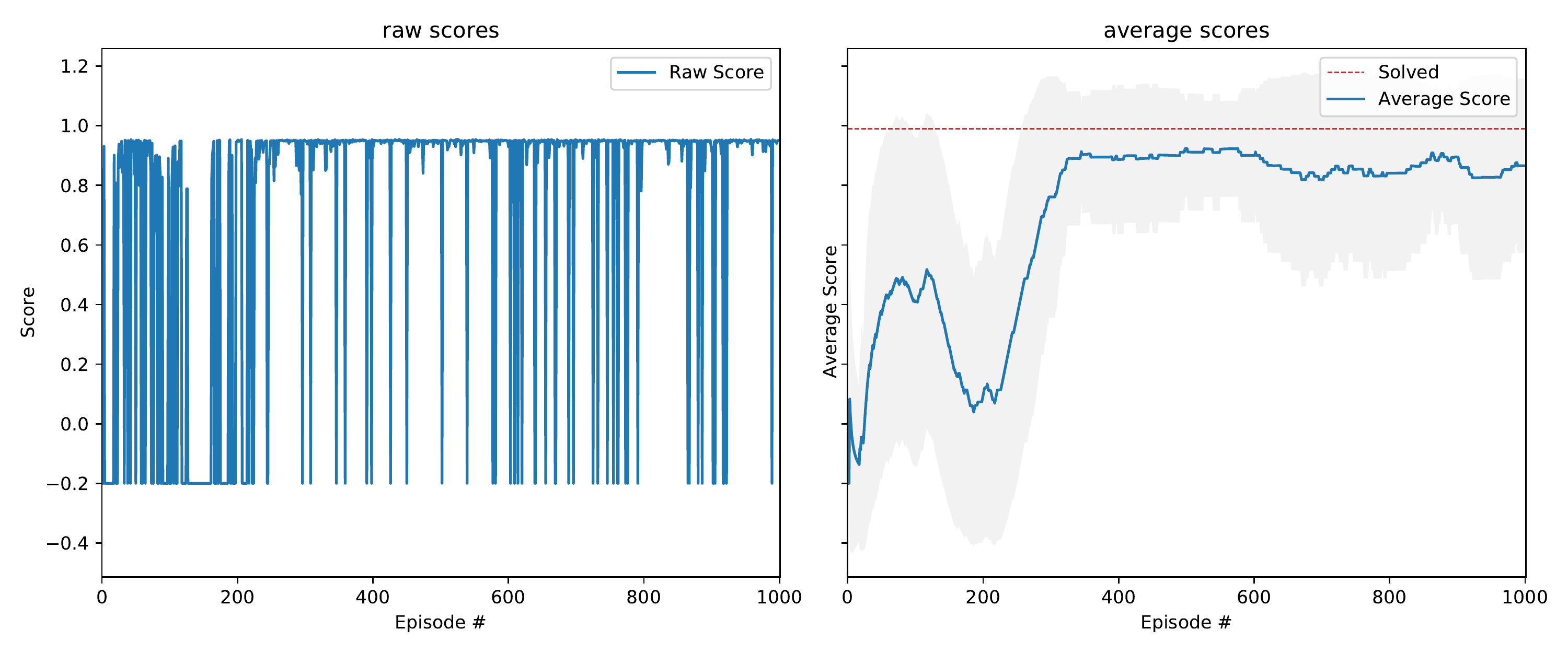}
            }
        \caption{\label{fig:noisy_noise001_X_fidelity_threshold_095} {\bfseries Training-from-scratch simulation for  \texttt{Environment-1}.}
        }
     \end{subfigure}
     \hfill
     \begin{subfigure}[t]{1\textwidth}
     \centering
         \scalebox{0.5}{
            \includegraphics{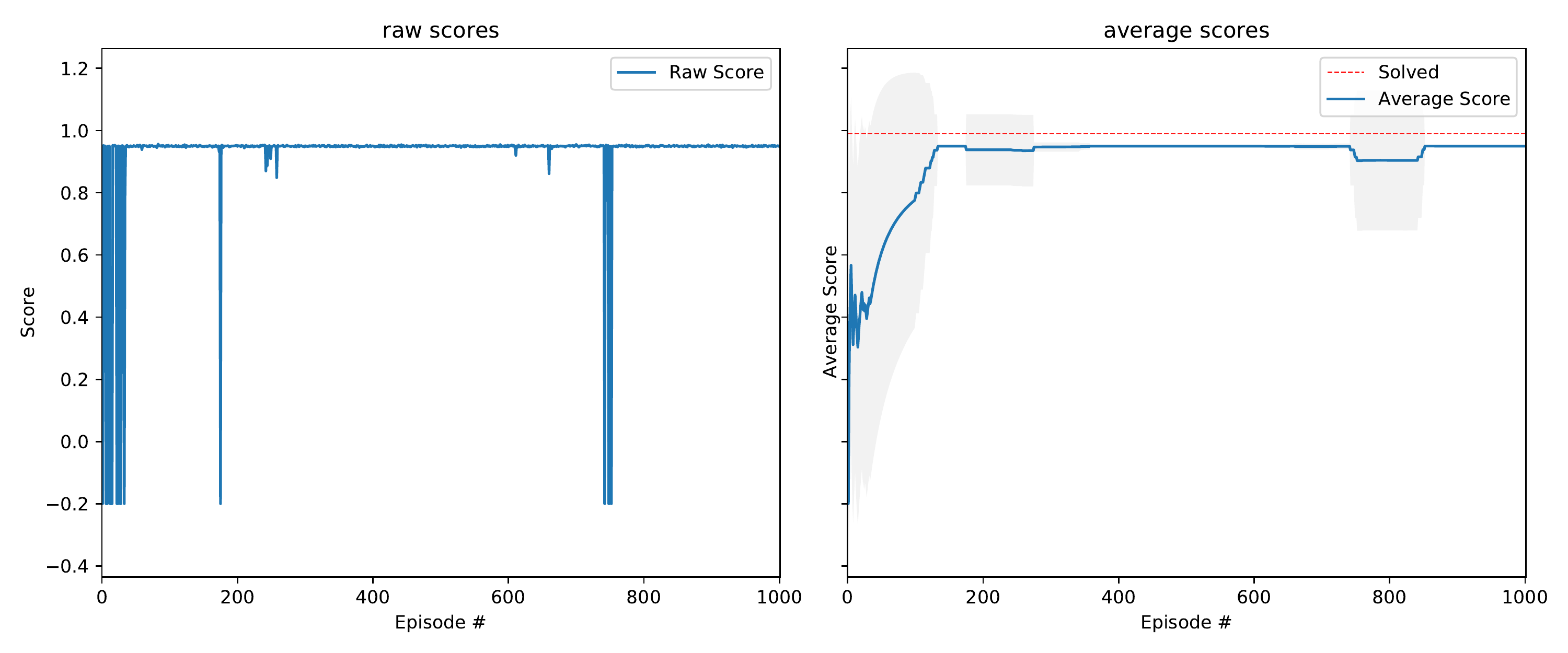}% Here is how to import EPS art
            }
        \caption{\label{fig:noisy_two_qubits_reuse_noise001_X_fidelity_threshold_095} {\bfseries Policy reuse simulation result for  \texttt{Environment-1}.} \\ \emph{Starting Policy Library: Simulation from  \texttt{Environment-0}.}}
     \end{subfigure}
     \hfill
    \caption{{\bfseries Training-from-scratch and policy reuse results for \texttt{Environment-1}:} Noisy two-qubit system with single-qubit error rate $0.01$ on the $X$ gate and fidelity threshold $0.95$.}
    \label{fig:env1}
\end{figure}

Now that we have the policy from \texttt{Environment-0} stored in our policy library, we can compare the two simulations involving \texttt{Environment-1}. The first is the training-from-scratch which shows how the agent performs under the environment with a regular DQN. Since the $X$-gate isn't required in the circuit solution for the Bell State, it is not very surprising that the number of episodes to solve this noisy environment is around $400$ (which is less than the previous noise-free environment simply due to the stochastic nature of the agent). The note-worthy difference is between the training-from-scratch (\figureautorefname{\ref{fig:noisy_noise001_X_fidelity_threshold_095}}) and the result using the PPR-algorithm (\figureautorefname{\ref{fig:noisy_two_qubits_reuse_noise001_X_fidelity_threshold_095}}). We can see that the PPR-algorithm converges to the correct circuit solution considerably earlier (around $200$ episodes in) than in the training-from-scratch simulation.

% \newpage

\subsection{Environment 2: $0.01$ Noise on the $X$ \& $H$ Gates}
\begin{figure}[!hbp]
     \begin{subfigure}[t]{1\textwidth}
     \centering
         \scalebox{0.5}{
            \includegraphics{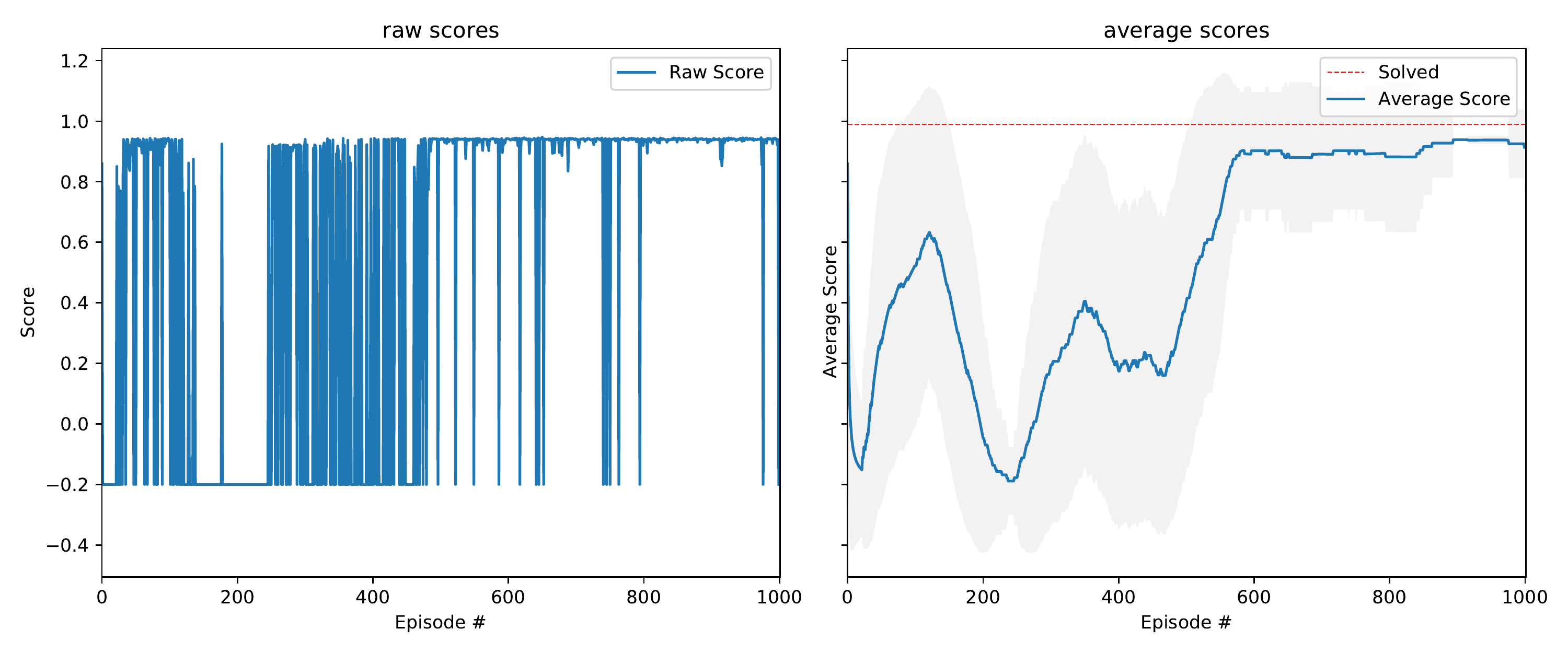}
            }
        \caption{\label{fig:noisy_noise001_X_H_fidelity_threshold_095} {\bfseries Training-from-scratch simulation for \texttt{Environment-2}.}
        }
     \end{subfigure}
     \hfill
     \begin{subfigure}[t]{1\textwidth}
     \centering
         \scalebox{0.5}{
            \includegraphics{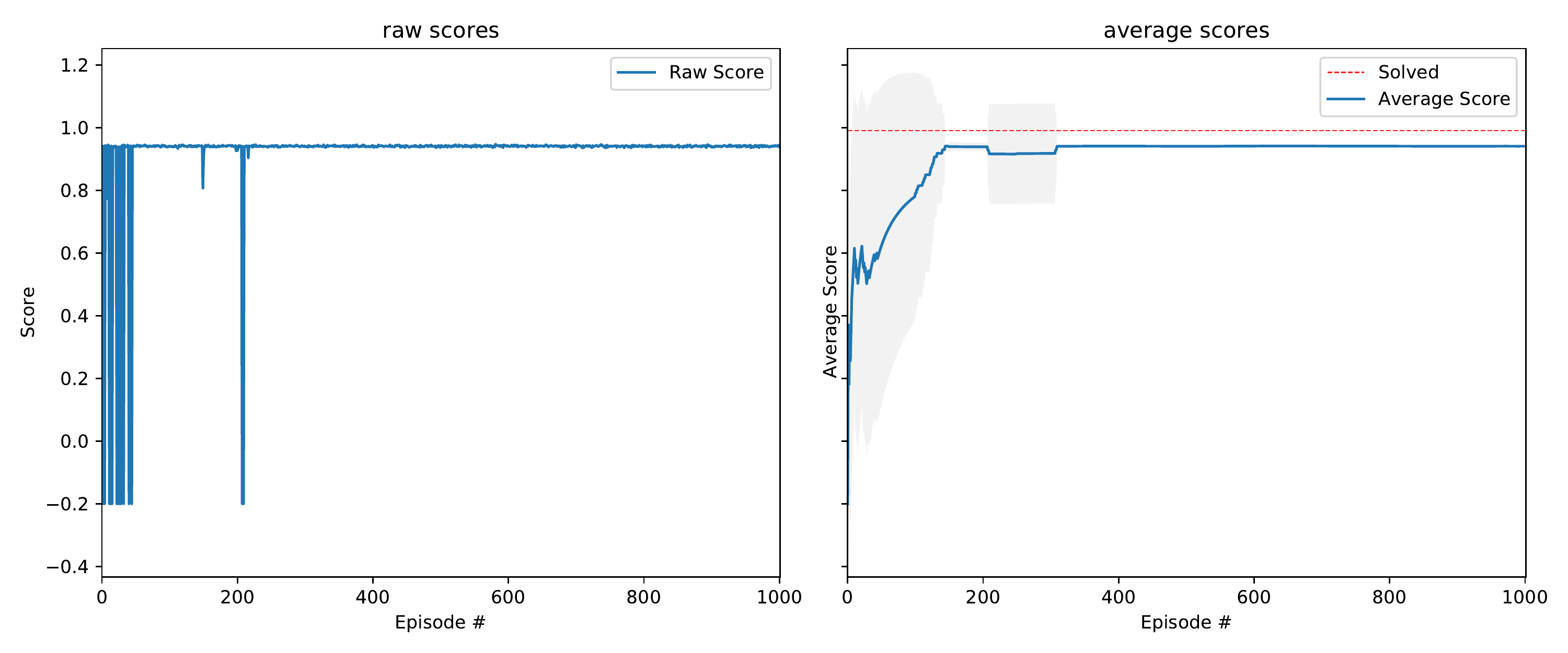}% Here is how to import EPS art
            }
        \caption{\label{fig:noisy_two_qubits_reuse_noise001_X_H_fidelity_threshold_095} {\bfseries Policy reuse simulation result for \texttt{Environment-2}.} \\ \emph{Starting Policy Library: from Scratch - \texttt{Environment-0}, Policy Reuse - \texttt{Environment-1}.}}
     \end{subfigure}
     \hfill
    \caption{{\bfseries Training-from-scratch and policy reuse results for \texttt{Environment-2}:} Noisy two-qubit system with single-qubit error rate $0.01$ on the $X$ \& $H$ gates and fidelity threshold $0.95$.}
    \label{fig:env2}
\end{figure}

Similarly for \texttt{Environment-2}, we see that the training-from-scratch (\figureautorefname{\ref{fig:noisy_noise001_X_H_fidelity_threshold_095}}) had a more difficult time than in the previous two environments because there is now an added noise on one of the necessary components of the circuit solution. In the training-from-scratch, the agent converges to the correct solution around the $600^{th}$ episode while with the PPR-algorithm (\figureautorefname{\ref{fig:noisy_two_qubits_reuse_noise001_X_H_fidelity_threshold_095}}) the agent reaches the right circuit in less than $200$ episodes. We can also see that even after reaching the final solution, the training-from-scratch (\figureautorefname{\ref{fig:noisy_noise001_X_H_fidelity_threshold_095}}) is still fairly unstable whereas the policy reuse result (\figureautorefname{\ref{fig:noisy_two_qubits_reuse_noise001_X_H_fidelity_threshold_095}}) converges and remains stable for the duration of the simulation. This makes the improvement with the PPR-algorithm even more apparent because it has taken a similar amount of time to solve what is a noticeably more challenging problem for the regular DQN learning scheme.

% \newpage

\subsection{Environment 3: $0.01$ Noise on the $X$ \& $CNOT$ Gates}
\begin{figure}[!htbp]
     \begin{subfigure}[t]{1\textwidth}
     \centering
         \scalebox{0.50}{
            \includegraphics{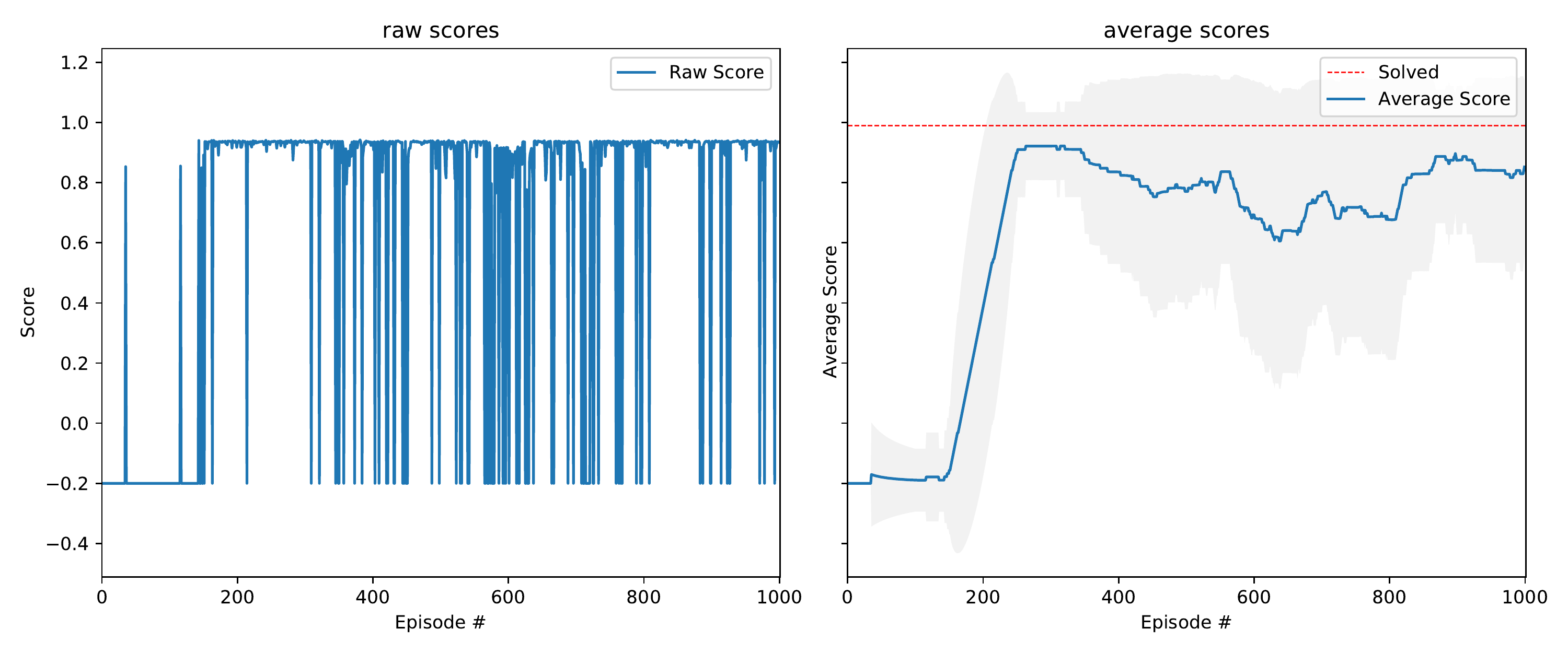}
            }
        \caption{\label{fig:noisy_noise001_X_CNOT_fidelity_threshold_095} {\bfseries Training-from-scratch simulation for \texttt{Environment-3}.}
        }
     \end{subfigure}
     \hfill
     \begin{subfigure}[t]{1\textwidth}
     \centering
         \scalebox{0.50}{
            \includegraphics{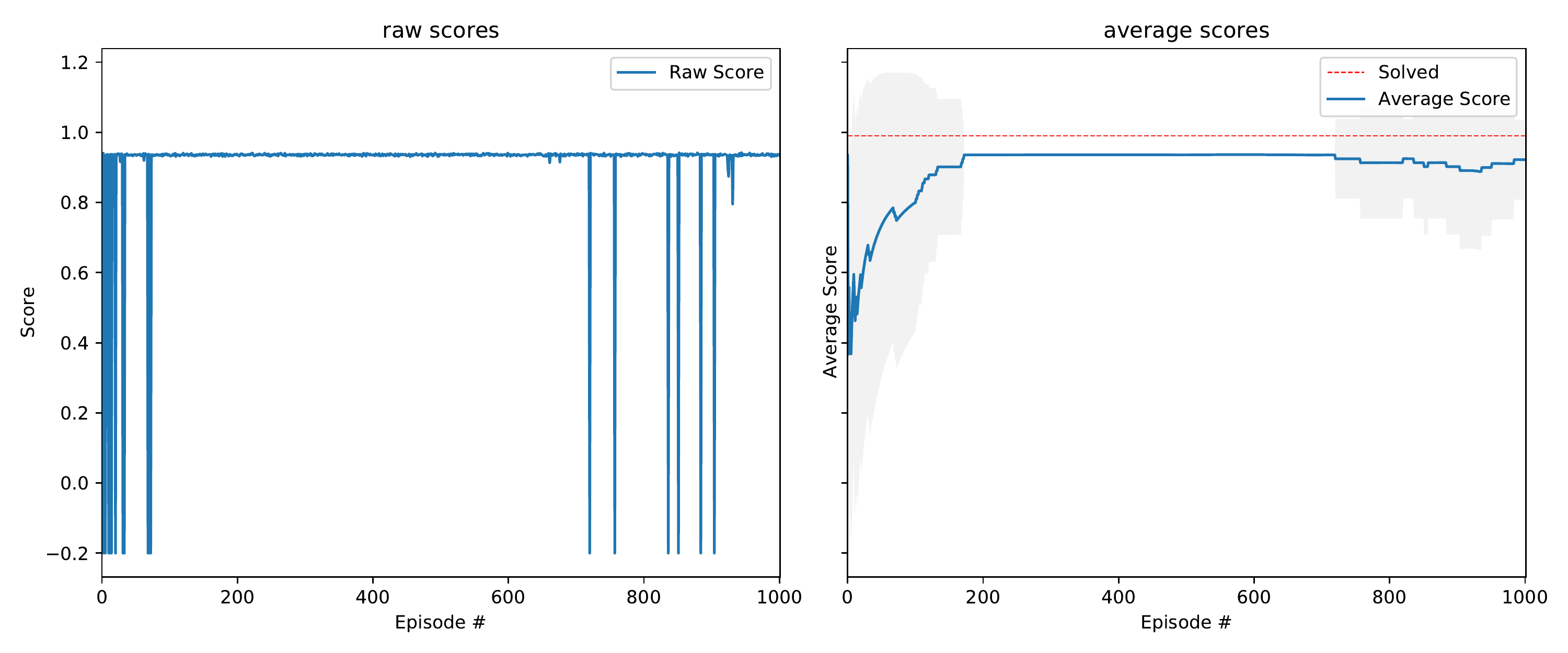}% Here is how to import EPS art
            }
        \caption{\label{fig:noisy_two_qubits_reuse_noise001_X_CNOT_fidelity_threshold_095} {\bfseries Policy reuse simulation result for \texttt{Environment-3}.} \\ \emph{Starting Policy Library: from Scratch - \texttt{Environment-0}, Policy Reuse - \texttt{Environment-1}, Policy Reuse - \texttt{Environment-2}.}}
     \end{subfigure}
     \hfill
    \caption{{\bfseries Training-from-scratch and policy reuse results for \texttt{Environment-3}:} Noisy two-qubit system with error rate $0.01$ on the $X$ \& $CNOT$ gates and fidelity threshold $0.95$.}
    \label{fig:env3}
\end{figure}

The next noise level involves the $CNOT$-gate, which affects both qubits and for that reason we expect it to be more challenging than the previous environments. We see that the training-from-scratch (\figureautorefname{\ref{fig:noisy_noise001_X_CNOT_fidelity_threshold_095}}) struggled to stabilize more with this added noise than in the previous environments, reaching the solution briefly around the $200^{th}$ episode but only stably getting the correct answer after episode $800$. Once again this highlights the performance of the PPR-algorithm (\figureautorefname{\ref{fig:noisy_two_qubits_reuse_noise001_X_CNOT_fidelity_threshold_095}}) since the agent again reaches the right circuit in less than $200$ episodes. 

\subsection{Environment 4: $0.005$ Noise on the $X$, $H$, \& $CNOT$ Gates}
\begin{figure}[!hbp]
     \begin{subfigure}[t]{1\textwidth}
     \centering
         \scalebox{0.50}{
            \includegraphics{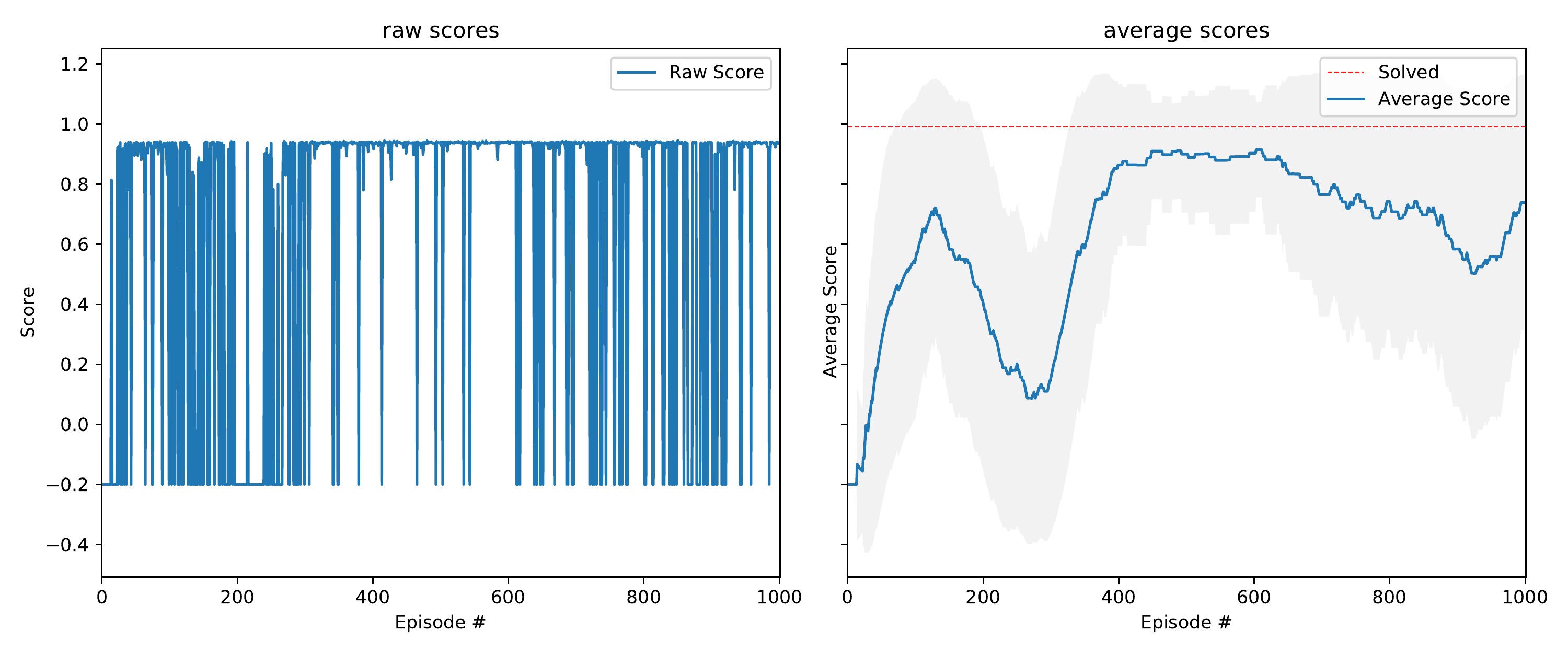}
            }
        \caption{\label{fig:noisy_noise0005_X_H_CNOT_fidelity_threshold_095} {\bfseries Training-from-scratch simulation for \texttt{Environment-4}.}
        }
     \end{subfigure}
     \hfill
     \begin{subfigure}[t]{1\textwidth}
     \centering
         \scalebox{0.50}{
            \includegraphics{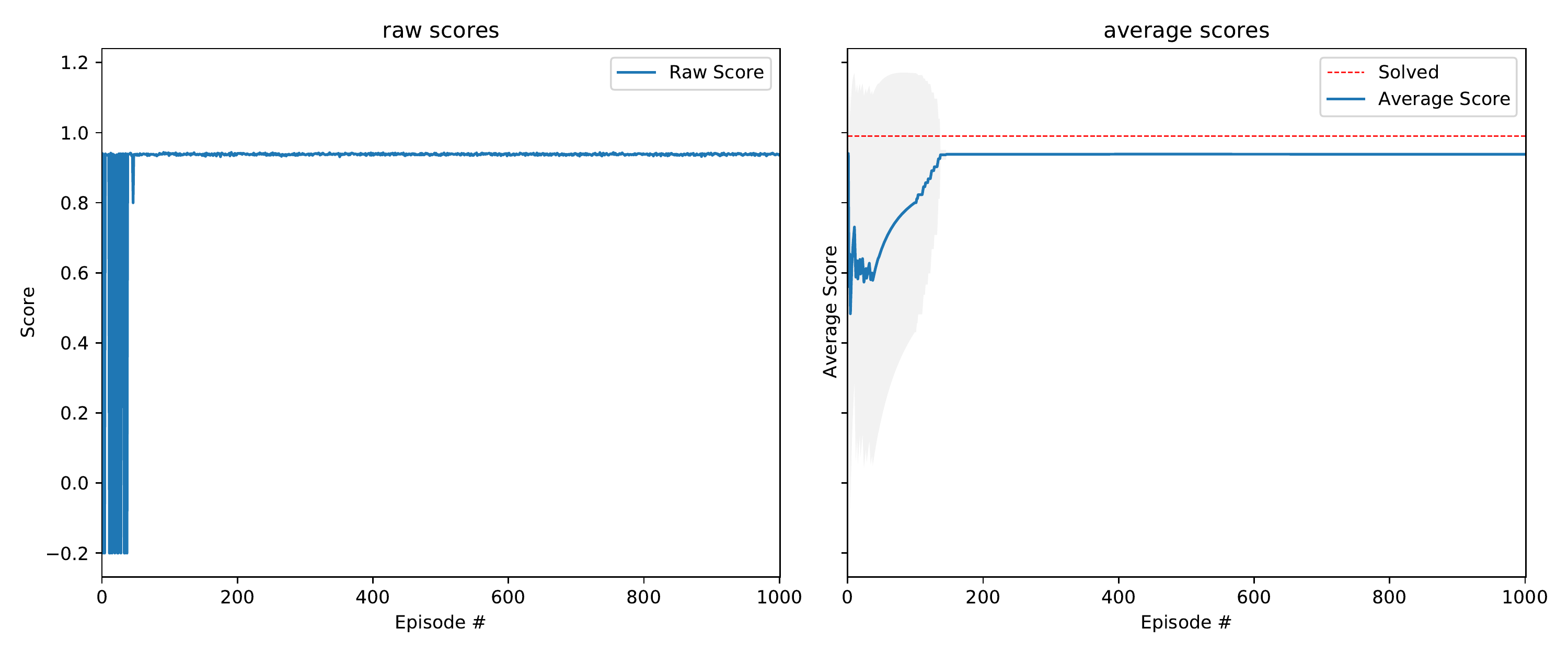}% Here is how to import EPS art
            }
        \caption{\label{fig:noisy_two_qubits_reuse_noise0005_X_H_CNOT_fidelity_threshold_095} {\bfseries Policy reuse simulation result for \texttt{Environment-4}.} \\ \emph{Starting Policy Library: from Scratch - \texttt{Environment-0}, Policy Reuse - \texttt{Environment-1}, Policy Reuse - \texttt{Environment-2}, Policy Reuse - \texttt{Environment-3}.}}
     \end{subfigure}
     \hfill
    \caption{{\bfseries Training-from-scratch and policy reuse results for \texttt{Environment-4}:} Noisy two-qubit system with error rate $0.005$ on the $X$, $H$, \& $CNOT$ gates and fidelity threshold $0.95$.}
    \label{fig:env4}
\end{figure}

The noise level for this environment involves adding noise on three gates, which is more challenging than only two gates as we had in the previous environments. We see that the training-from-scratch (\figureautorefname{\ref{fig:noisy_noise0005_X_H_CNOT_fidelity_threshold_095}}) again struggled to stabilize with this added noise but unlike the previous environment, the agent doesn't appear to converge to the optimal solution within the 1000 episode time frame of the simulation. Conversely, we can see that the PPR-algorithm (\figureautorefname{\ref{fig:noisy_two_qubits_reuse_noise0005_X_H_CNOT_fidelity_threshold_095}}) again reaches the right circuit in less than $200$ episodes and stabilizes very well. 

\subsection{Environment 5: $0.01$ Noise on the $X$ \& $H$ Gates, $0.005$ on the $CNOT$ Gate}
\begin{figure}[!htbp]
     \begin{subfigure}[t]{1\textwidth}
     \centering
         \scalebox{0.50}{
            \includegraphics{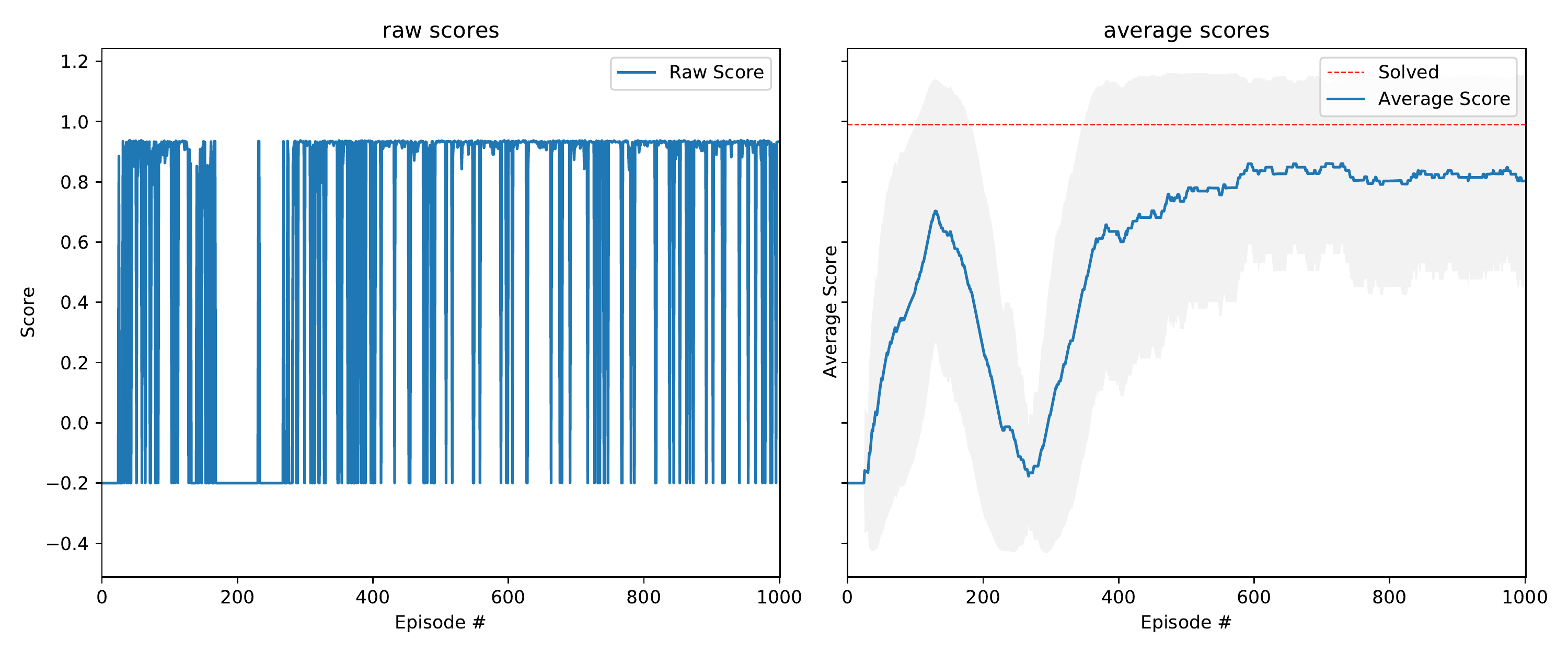}
            }
        \caption{\label{fig:noisy_noise0010010005_X_H_CNOT_fidelity_threshold_095} {\bfseries Training-from-scratch simulation for \texttt{Environment-5}.}
        }
     \end{subfigure}
     \hfill
     \begin{subfigure}[t]{1\textwidth}
     \centering
         \scalebox{0.50}{
            \includegraphics{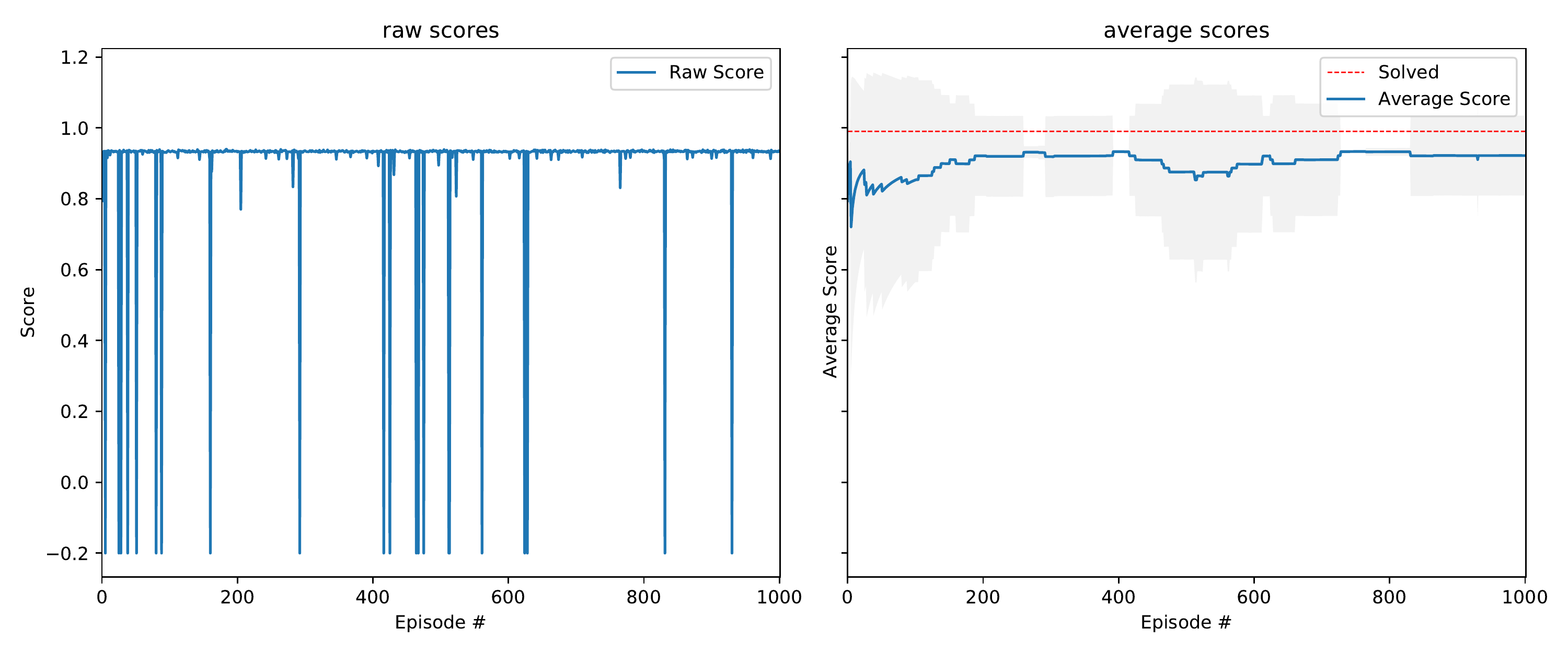}% Here is how to import EPS art
            }
        \caption{\label{fig:noisy_two_qubits_reuse_noise0010010005_X_H_CNOT_fidelity_threshold_095} {\bfseries Policy reuse simulation result for \texttt{Environment-5}.} \\ \emph{Starting Policy Library: from Scratch - \texttt{Environment-0}, Policy Reuse - \texttt{Environment-1}, Policy Reuse - \texttt{Environment-2}, Policy Reuse - \texttt{Environment-3}, Policy Reuse - \texttt{Environment-4}.}}
     \end{subfigure}
     \hfill
    \caption{{\bfseries Training-from-scratch and policy reuse results for \texttt{Environment-5}:} Noisy two-qubit system with error rate $0.01$ on the $X$ \& $H$ gates, error rate $0.005$ on the $CNOT$ gate, and fidelity threshold $0.95$.}
    \label{fig:env5}
\end{figure}

The noise level for this environment increases the amount of noise on two of the three gates, specifically the $X$ and $H$ gates, making it more difficult than \texttt{Environment-4}. We can see that the training-from-scratch (\figureautorefname{\ref{fig:noisy_noise0010010005_X_H_CNOT_fidelity_threshold_095}}) is noticeably more unstable (dipping from $0.7$ to nearly $-0.2$ between episode $150$ to $300$). From the results of the PPR-algorithm (\figureautorefname{\ref{fig:noisy_two_qubits_reuse_noise0010010005_X_H_CNOT_fidelity_threshold_095}}) we see that with the policy reuse, the agent performs much better and although there is some variation from the optimal solution, it is still a drastic improvement compared to the training-from-scratch (\figureautorefname{\ref{fig:noisy_noise0010010005_X_H_CNOT_fidelity_threshold_095}}) and this deviation is understandable due to the difficulty of the noise in this environment.

% \par\vfill\break
% \advance\vsize by -2cm % Return old margins and page height
% \advance\voffset by -2cm % Return old margins and page height
% \newpage
\section{\label{sec:Discussion}Discussion}
\subsection{Relevant Works}
%\YC{Discuss different ML-based QAS, QECC, Quantum Control work}
%\EY{Mix of sources that cover RL for QC and the QAS objective in particular: Distinguishing the specialty of this work compared to the others. (Novelty)}
%
%\EY{sources just added}
AI and ML techniques have been used to tackle certain challenges in quantum computing. Notable examples are quantum architecture search \cite{chen2021quantum}, quantum error correction \cite{chen2019machine, convy2021machine} and quantum control \cite{zeng2020quantum}. Among various ML techniques, RL is the one which draws a lot of attention since it is designed to solve complex sequential decision making problems and indeed, it has been applied in several recent works \cite{nautrup2019optimizing,andreasson2019quantum,colomer2020reinforcement,sweke2020reinforcement}. 
RL has been used to optimize existing quantum circuit architectures \cite{fosel2021quantum,ostaszewski2021reinforcement}. The idea is that there is a given ansatz of which the performance may not be optimal. The RL is used to modify the architecture by dropping some of the components or fine-tuning the circuit parameters \cite{fosel2021quantum,ostaszewski2021reinforcement}. The same logic is applied to optimizing existing quantum error correction codes \cite{nautrup2019optimizing,andreasson2019quantum,colomer2020reinforcement,sweke2020reinforcement}.

Our approach in this paper is different-- we do not have an ansatz for the RL agent to optimize. On the contrary, the RL agent is trained to generate the quantum gate sequence from scratch, given only fidelity values and quantum measurement values. This task is expected to be harder than finding some potential modifications from existing non-optimal architectures. Indeed, the previous work \cite{kuo2021quantum} considers this direction, providing a RL framework to find quantum architectures from scratch. However, for different tasks with various noise patterns, the RL agents need to be trained from scratch as well, making the workflow inefficient for constantly changing or drifting device noises and subsequently motivating our implementation of continual learning to address this problem.

\subsection{Scalability}
\subsubsection{Policy Library}
In this framework, the Policy Library grows with each run of the algorithm since each new task will be adding a new policy after completion. This becomes an issue, particularly for storage, as the number of tasks grows to be very large. With a large number of tasks, we would expect the Policy Library to contain a large number of policies and, in addition to storage, this effects the efficiency of the algorithm since it will need to sample from the large library set. 

A possible solution to this is the use of \emph{eigenpolicies}, which was proposed in the Fernandez paper \cite{fernandez2006probabilistic}. Eigenpolicies operate on the principle that new policies which have a high similarity to a previous policy in the library will be absorbed by a principal eigenpolicy rather than occupy its own separate spot within the library. 

In implementing the creation and maintaining of these eigenpolicies, the Policy Library can condense multiple policies by choosing whether it should store the newly generated policy as the start of a new eigenpolicy or absorb it into a previous entry. 

Fascinatingly, this addition to the quantum circuit construction framework would allow for the existence of base circuit compositions. If we think of a classical computing analogy, the eigenpolicies would correspond to circuit blocks that can be readily reused to compose more complex, and larger-scale circuits. In other words, the eigenpolicies will correspond to a network of weights for a particular quantum state and if it is unique enough from other states, it will remain an eigenpolicy in the library and be used for problems that require it.

\subsubsection{Action Space}
%\EY{Use N comparison; note for self-- information bottleneck.}
%\EY{Include from previous paper}
In the framework detailed within this paper, the action space of the DQN (described in \equationautorefname{\ref{eq:action gates}}) is coded such that each qubit has node for each gate in the set of given gates. This implies that as the number of qubits grows, the dimension of the action space grows non-linearly (due to the combinations of the $CNOT$ gates). For the general $n$-qubit case, there will be $5n+n(n-1)= \Omega(n^2)$ actions. This makes scaling to a system with a large number of qubits fairly strenuous given that the action space is quadratic with respect to $n$. To counter this issue, a possible course of action is to have another level of network usage (one to generate the qubit index --which scales linearly-- and the other to generate gates).

\subsection{Training on a Real Quantum Computer}
%\EY{Reference current noise error for state of the art QCs and justify why we chose 0.01 and point to future directions. This paper simplifies the concept of qc errors and can be used as a starting point to work with more complex errors in the future.}
%\EY{cite white papers on QC resources; Point out that cloud-based QC resources are not as developed but in the event it improves the framework will be able to train real QCs correctly; mention queue and difficulties for RL which requires many runs. Identify it's an engineering problem, approximate a time-frame for achieving goal.}
A necessary consideration for any research with simulated quantum computing is how well the work will translate to a real quantum computer. The current state of the quantum computers impose a realistic limitation on the number of gates that can be used on any given qubit due to the instability of the hardware and specifically, the severity of the error, which accumulates as more gates are added.

The work presented in this paper is designed to accommodate the IBM-Q interface and it would be trivial to apply the framework onto a real quantum computer. The more pressing concern with this decision lies in the state of the cloud-based facility. This includes the challenge of dealing with the unaltered device noise (as mentioned in \sectionautorefname{\ref{sec:Noise}}) as well as the accessibility to the program while it is running. For RL in particular, many runs are required and these runs must be performed sequentially which can be challenging for a platform designed to be shared (in which jobs are queued). This, however, presents itself as more of a logistical or engineering problem and efforts are currently being made to develop tools which allow users more easily run experiments with many iterations on these cloud-based QC platforms (e.g. Qiskit Runtime). 
\subsection{Continual Learning with Policy Gradient Algorithms}
%\YC{Discuss the potential of using continual learning in other policy gradient RL}
%\YC{Update the citation  / ref for papers using Policy Gradient for quantum computing}
%\EY{mention previous policy grad in prev paper is very successful but the implementation of CL for those algorithms hasn't been explored yet, but is hopeful}

A family of RL algorithms based on policy gradient methods are highly successful nowadays \cite{mnih2016asynchronous}. Previously, it has been demonstrated that policy gradient algorithms can successfully implement the quantum architecture search objectives \cite{kuo2021quantum}, quantum compiling \cite{herrera2021policy, moro2021quantum} and quantum control tasks \cite{baum2021experimental}. One of the major challenges in training policy gradient RL agents is the sampling of a large amount of different paths (rollout), which requires significant computational resources. Therefore, there is a very strong motivation for continual learning to be implemented in those algorithms such that their usage can be optimized and further reduce the resource requirement of the problem. Continual RL in the context of policy gradients have been studied in works \cite{kim2021policy} under standard RL tasks. It is interesting to study the potential of such algorithms in the challenging QAS tasks. We leave this for future investigation.

\section{\label{sec:Conclusion}Conclusion}
%\EY{Reiterate contribution and benefits of work for QC community; refer to other papers (should be more concise and move out some information that belongs in the discussion section}

%\EY{Condense the conclusion.}
%\YC{I condense the conclusion. Please check.}
%From the work shown in this paper, we identify two areas that can be developed upon. The first is the subject of noise. Although the PPR model can handle a noise level of 0.01 as explained in the \sec, it is still not powerful enough to solve with the real noise taken from the IBM devices. A possible direction for future work is to look into how to accommodate that noise level. The second direction is the consideration for scaling. As mentioned in the DQN subsection, the number of actions would grow linearly with the number of qubits of the system given the distinction between gates for separate qubits. This may pose as a problem as the DQN would need to deal with a larger dimensionality in its action space and subsequently take longer to compute.

% Challenge, methodology, results and impact

In this paper, we provide a framework to study how continual reinforcement learning can be applied to QAS problems. 
% We demonstrated how to build a quantum circuit step-by-step via continual reinforcement learning methods, which overcomes the obstacle of needing to retrain the agent for environments with different noise patterns. 
%results%
% We showed that using the PPR-algorithm lets the agent make the right circuit in a shorter amount of time if the task given to it has a high similarity to one that it has solved and remembered previously. 
We showed numerically that the deep Q-learning with probabilistic policy reuse (PPR) lets the RL agent learn a good policy for the construction of a quantum gate sequence in various unseen environments. Furthermore, we demonstrate that the algorithm accomplishes this task within a significantly reduced number of training episodes through the leveraging of previously learned policies. 
%Impact%
The proposed framework is fairly general and can be applied to different types of quantum computing challenges. The result is especially useful when the underlying noise of the system is not well-understood. This in turn provides a promising implication to counter noise in quantum computers by enabling real-time autonomous corrections to the system in noisy environments in order to reach an arbitrary desired output state. 
 
% Overall, the machine learning framework poses as a strong candidate in the area of quantum error correction and other subsidiary probabilistic challenges that quantum computers face. As this paper demonstrated, the usage of classical computing techniques may be the key to stabilizing and improving these complex quantum systems in the future.

\begin{acknowledgments}
We thank Meifeng Lin, Tzu-Chieh Wei, Leo Fang, and Elisha Siddiqui for the fruitful discussions.
This work is supported by the U.S.\ Department of Energy, Office of Science, Office of High Energy Physics program under Award Number DE-SC-0012704, Office of Workforce Development for Teachers and Scientists (WDTS) under the Science Undergraduate Laboratory Internships Program (SULI) and the Brookhaven National Laboratory LDRD \#20-024. 

\end{acknowledgments}
\par\vfill\break

\appendix
% \EY{Make sure the latest version of the code is here}
\section{\label{sec:Appendix}Appendixes}
%\par\vfill\break % Break Last Page
\advance\vsize by 2cm % Advance page height
\advance\voffset by -1cm % Shift top margin
\begin{algorithm}[H]
\begin{algorithmic}
\State \textbf{Define} the transition pair to be in the form $(s_{t}, a_{t}, r_{t}, s_{t+1})$
\State \textbf{Given} an environment, $\mathcal{E}$
\State \textbf{Given} the replay memory, $\mathcal{D}$
\State \textbf{Given} a new policy, $L_0$
\State \textbf{Given} the maximum number of episodes to execute, $K$
\State \textbf{Given} the maximum number of steps per episode, $H$
\State \textbf{Initialize} the rewards list $W$
\For{episode $=1,2,\ldots,K$} 
\State Reset the environment $\mathcal{E}$
    \For{$t = 1,2,\ldots,H$}
    \State Select action $a_{t}$ from $L_0$ using greedy policy selection
    \State Perform the action $a_{t}$
    \State Observe the new state $s_{t +1}$
    \State \textbf{if} not done:
        \State \quad Assign the next state in the transition pair to be the new state $s_{t+1}$
    \State \textbf{else}:
        \State \quad Assign the next state in the transition pair to be none
    \State Store the transition pair to $\mathcal{D}$ and assign current state $s$ to be next state $s_{t+1}$
    \State Optimize $L_0$
    \State \textbf{if} done:
    \State \quad Update $W$
    \State \quad break
\EndFor
\EndFor
\State \textbf{Return} mean($W$), new policy $L_0$
\end{algorithmic}
\caption{q-learning algorithm}
\label{q_learn_alg}
\end{algorithm}
\par\vfill\break
\advance\vsize by -2cm % Return old margins and page height
\advance\voffset by 1cm % Return old margins and page height

\par\vfill\break % Break Last Page
\advance\vsize by 2cm % Advance page height
\advance\voffset by -1cm % Shift top margin
\scalebox{0.9}{
\begin{minipage}{\linewidth}
\begin{algorithm}[H]
\begin{algorithmic}
\State \textbf{Define} the transition pair to be in the form $(s_{t}, a_{t}, r_{t}, s_{t+1})$
\State \textbf{Given} an environment, $\mathcal{E}$
\State \textbf{Given} the replay memory, $\mathcal{D}$
\State \textbf{Given} a past policy to try $L_i$ and the new policy $L_0$
\State \textbf{Given} the maximum number of episodes to execute, $K$
\State \textbf{Given} the maximum number of steps per episode, $H$
\State \textbf{Given} the parameters $\psi$ and $\nu$
\State \textbf{Initialize} the rewards list $W$
\For{episode $=1,2,\ldots,K$} 
\State Reset the environment $\mathcal{E}$
    \For{$t$ $=1,2,\ldots,H$}
    \State Sample a random probability $p$
    \State \textbf{if} $p <= \psi$
    \State \quad Select action $a_{t}$ from $L_i$ using greedy policy selection
    \State \textbf{else}:
    \State \quad Select action $a_{t}$ from $L_0$ using greedy policy selection
    \State Perform the action $a_{t}$
    \State Observe the new state $s_{t+1}$
    \State Update $\psi$ to be $\psi \cdot \nu$
    \State \textbf{if} not done:
        \State \quad Assign the next state in the transition pair to be the new state $s_{t+1}$
    \State \textbf{else}:
        \State \quad Assign the next state in the transition pair to be none
    \State Store the transition pair to $\mathcal{D}$ and assign current state to be next state
    \State Optimize $L_0$
    \State \textbf{if} done:
    \State \quad Update $W$
    \State \quad break
\EndFor
\EndFor
\State \textbf{Return} mean($W$), new policy $L_0$
\end{algorithmic}
\caption{$\pi$-exploration algorithm}
\label{policy_reuse_pi_alg}
\end{algorithm}
\end{minipage}
}
\par\vfill\break
\advance\vsize by -2cm % Return old margins and page height
\advance\voffset by 1cm % Return old margins and page height

\par\vfill\break % Break Last Page
\advance\vsize by 2cm % Advance page height
\advance\voffset by -1cm % Shift top margin
\scalebox{0.9}{
\begin{minipage}{\linewidth}
\begin{algorithm}[H]
\begin{algorithmic}
\State \textbf{Given} a new task $\Omega$ we want to solve
\State \textbf{Given} a policy library $L = \{\Pi_{1}, \cdots, \Pi_{n} \}$
\State \textbf{Given} an initial temperature parameter $\tau$ and the incremental size $\delta\tau$ for the Boltzmann policy selection strategy
\State \textbf{Given} the maximum number of episodes to execute, $K$
\State \textbf{Given} the maximum number of steps per episode, $H$
\State \textbf{Given} the parameters $\psi$ and $\nu$ for the $\pi$-exploration strategy
\State \textbf{Initialize} replay memory $\mathcal{D}$ to capacity $N$
\State \textbf{Initialize} a new policy $L_{0}$
\State \textbf{Initialize} a target policy $\hat{L}_{0} = L_{0}$
\State \textbf{Initialize} target network update rate, $T$  
\State \textbf{Initialize} the rewards $W_{i}$ to $0$
\State \textbf{Initialize} the number of episodes where policy $\Pi_{i}$ has been chosen, $U_{i} = 0, \forall i = 1, \cdots, n$
\For{episode $=1,2,\ldots,K$} 
\State Get probability vector $p$ from softmax function inputting $W$ and $\tau$
\State Sample from $p$ to get an index for the action policy $k$
    \State \textbf{if} {$k == 0$}:
        \State \quad Use the q-learning algorithm to get rewards $R$ and $L_0$
    \State \textbf{else}:
        \State \quad Use the $\pi$-exploration algorithm to get $R,\ L_0$
    \State Update $W_{k} \gets \frac{W_{k} \cdot U_{k} + R}{U_{k} + 1}$ 
    \State Update $U_{k} \gets U_{k} + 1$
    \State Update $\tau \gets \tau + \delta\tau$
    \State \textbf{if} episode is divisible by $T$:
        \State \quad Update $\hat{L}_{0}$ with $L_0$
\EndFor
\end{algorithmic}
\caption{Probabilistic Policy Reuse with deep Q learning}
\label{policy_reuse_alg}
\end{algorithm}
\end{minipage}
}
\par\vfill\break
\advance\vsize by -2cm % Return old margins and page height
\advance\voffset by 1cm % Return old margins and page height

\bibliographystyle{ieeetr}
\bibliography{bib/ml,bib/rl_qcomp,bib/ml_qas,bib/ml_qec,bib/ml_qctrl,bib/packages,bib/pr,bib/sam,bib/qec,bib/qc,bib/nisq,bib/vqc,bib/qml_examples,bib/rl,bib/rl_quantum_control,bib/rl_quantum_error_correction,bib/relevant_qas,bib/continual_learning,bib/classical_nas,bib/rl_qas}% Produces the bibliography via BibTeX.

\end{document}